# A General Framework for Portfolio Theory. Part I: theory and various models

Stanislaus Maier-Paape[*]     Qiji Jim Zhu[†]

October 11, 2017

**Abstract**

Utility and risk are two often competing measurements on the investment success. We show that efficient trade-off between these two measurements for investment portfolios happens, in general, on a convex curve in the two dimensional space of utility and risk. This is a rather general pattern. The modern portfolio theory of Markowitz [15] and its natural generalization the capital market pricing model [22] are special cases of our general framework when the risk measure is taken to be the standard deviation and the utility function is the identity mapping. Using our general framework we also recover the results in [20] that extends the capital market pricing model to allow for the use of more general deviation measures. This generalized capital asset pricing model also applies to e.g. when an approximation of the maximum drawdown is considered as a risk measure. Furthermore, the consideration of a general utility function allows to go beyond the "additive" performance measure to a "multiplicative" one of cumulative returns by using the log utility. As a result, the growth optimal portfolio theory [9] and the leverage space portfolio theory [28] can also be understood under our general framework. Thus, this general framework allows a unification of several important existing portfolio theories and goes much beyond.

**Acknowledgement.** We thank Andreas Platen for his constructive suggestions after reading earlier versions of the manuscript.

[*]Institut für Mathematik, RWTH Aachen University, 52062 Aachen, Templergraben 55, Germany *maier@instmath.rwth-aachen.de*

[†]Department of Mathematics, Western Michigan University, 1903 West Michigan Avenue, Kalamazoo, MI 49008, *zhu@wmich.edu*

# 1 Introduction

The Markowitz modern portfolio theory [15] pioneered the quantitative analysis of financial economics. The most important idea proposed in this theory is that one should focus on the trade-off between expected return and the risk measured by the standard deviation. Mathematically, the modern portfolio theory leads to a quadratic optimization problem with linear constraints. Using this simple mathematical structure Markowitz gave a complete characterization of the efficient frontier for trade-off the return and risk. Tobin [26] showed that the efficient portfolios as an affine function of the expected return. Markowitz portfolio theory was later generalized by Lintner [9], Mossin [17], Sharpe [22] and Treynor [25] in the capital asset pricing model (CAPM) by involving a riskless bond. In the CAPM model, both the efficient frontier and the related efficient portfolios are affine in terms of the expected return [22, 26].

The nice structures of the solutions in the modern portfolio theory and the CAPM model afford many applications. For example, the CAPM model is designed to provide reasonable price for risky assets in the market place. Sharpe used the ratio of excess return to risk (called the Sharpe ratio) to provide a measurement for investment performance [23]. Also the affine structure of the efficient portfolio in terms of the expected return leads to the concept of a market portfolio as well as the two fund theorem [26] and the one fund theorem [22, 26]. These results provided a theoretical foundation for passive investment strategies.

While using the expected return and standard deviation as measures for reward and risk of a portfolio brings much convenience in the mathematical analysis, many other measures are more realistic. Since Bernoulli studied the St. Petersburg paradox [2], concave utility functions have been widely accepted as a more appropriate measure of the reward. General expected utilities have been used in many cases to measure the performance of a portfolio. On the other hand, current drawdown [13], maximum drawdown and its approximations [10, 12, 30], deviation measure [20], conditional value at risk [19] and more abstract coherent risk measures [1] are widely used as risk measures in practices. A common thread in these risk measures is that they are convex reflecting the belief that diversification reduces risk. The goal of this paper is to extend the modern portfolio theory into a general framework under which one can analyze efficient portfolios that trade-off between a convex risk measure and a reward captured by an expected utility. We phrase our primal problem as a convex portfolio optimization problem of minimizing a convex risk measure subject to the constraint that the expected utility of the portfolio is above a certain level. Thus, convex duality plays a crucial role and the structure of the solutions to both the primal and dual problems often have significant financial implications. We show that, in the space of risk measure and expected utility, efficient trade-off happens on an increasing concave curve. We also show that the efficient portfolios continuously depend on the level of the expected utility.

The Markowitz modern portfolio theory and the capital asset pricing model are, of course, special cases of this general theory. Markowitz determines portfolios of purely risky assets which provide an efficient trade-off between expected return and risk measured by the standard deviation (or equivalently the variance). Mathematically, this is a class of convex programming problems of minimizing the standard deviation of the portfolio parameterized by the level of the expected returns. The capital asset pricing model, in essence, extends the Markowitz modern portfolio theory by including a riskless bond in the portfolio. We observe that the space of the risk-expected return is, in fact, the space corresponding to the dual of the Markowitz portfolio problem. The shape of the famous Markowitz bullet is a manifestation of the well known fact that the optimal value function of a convex programming problem is convex with respect to the level of constraint. As mentioned above, the Markowitz portfolio problem is a quadratic optimization problem with linear constraint. This special structure of the problem dictates the affine structure of the optimal portfolio as a function of the expected return (see Theorem 4.1). This affine structure leads to the important two fund theorem that provides a theoretical foundation for the passive investment method. For the capital asset pricing model, such an affine structure appear in both the primal and dual representation of the solutions which leads to the two fund separation theorem in the portfolio space and the capital market line in the dual space of risk-return trade-off (cf. Theorem 4.5).

The flexibility in choosing different risk measures allows us to extend the analysis of the essentially quadratic risk measure pioneered by Markowitz to a wider range. For example, when the risk measure is a deviation measure [20], which happens e.g. when an approximation of the current drawdown is considered (see [14]), and the expected return is used to gauge the performance we show that the affine structure of the efficient solution in the classical capital market pricing model is preserved (cf. Theorem 5.1), recovering in particular the results in [20]. This is significant in that it shows that the passive investment strategy is justifiable in a wide range of settings.

The consideration of a general utility function, however, allows us to go beyond the "additive" performance measure in modern portfolio theory to a "multiplicative" one including cumulative returns when, for example, using the log utility. As a result the growth optimal portfolio theory [9] and the leverage space portfolio theory [28] can also be understood under our general framework. The optimal growth portfolio pursues to maximize the expected log utility which is equivalent to maximize the expected cumulative compound return. It is known that the growth optimal portfolio is usually too risky. Thus, practitioners often scale back the risky exposure from a growth optimal portfolio. In our general framework, we consider the portfolio that minimizes a risk measure given a fixed level of expected log utility. Under reasonable conditions, we show that such portfolios form a path parameterized by the level of expected log utility in the portfolio space that connects the optimal growth portfolio and the portfolio of a riskless bond (see Theorem 6.4). In general, for different risk measures we will derive different paths. These paths provide justifications for risk reducing curves proposed in the leverage space portfolio theory [28]. The dual problem projects the efficient trade-off path into a concave curve in the risk-expected log utility space parallel to the role of Markowitz bullet in the modern portfolio theory and the capital market line in the capital asset pricing model.

Unlike the modern portfolio theory and the capital asset pricing model, under the no arbitrage assumption, the efficient frontier here is usually a finite increasing concave curve. The lower left endpoint of the curve corresponds to the portfolio of pure riskless bond and the upper right endpoint corresponds to the growth optimal portfolio. The increasing nature of the curve tells us that the more risk we take the more cumulative return we can expect. The concavity of the curve indicates, however, that with the increase of the risk the marginal increase of the expected cumulative return will decrease. Thus, a risk averse investor will usually not choose the optimal growth portfolio. It is also interesting to observe that considering the dual problem corresponding to the growth optimal portfolio problem will leads to a version of the fundamental theorem of asset pricing (see Theorem 6.10) that connects the existence of an equivalent martingale measure to no arbitrage.

Besides unifying the several important results laid out above, the general framework has many new applications. In this first installment of the paper, we layout the framework, derive the theoretical results of crucial importance and illustrate them with a few examples. More new applications will appear in the subsequent papers [3, 14]. We arrange the paper as follows: First we discuss necessary preliminaries in the next section. Section 3 is devoted to our main result: a framework to trade-off between risk and utility of portfolios and its properties. In Section 4 we give a unified treatment of Markowitz portfolio theory, capital asset pricing model, and the Sharpe ratio. Section 5 is devoted to a discussion on the conditions under which the optimal trade-off portfolio possesses an affine structure. Section 6 discusses growth optimal portfolio theory and leverage portfolio theory. We also highlight some related important applications such as the fundamental theorem of asset pricing. We conclude in Section 7 pointing to applications worthy of further investigation.

## 2 Preliminaries

### 2.1 A portfolio model

We consider a simple one period financial market model $S$ on an economy with finite states represented by a sample space $\Omega = \{\omega_1, \omega_2, \ldots, \omega_N\}$. We use a probability space $(\Omega, 2^\Omega, P)$ to represent the states of the economy and their corresponding probability of occurring, where $2^\Omega$ is the algebra of all subsets of $\Omega$. The space of random variables on $(\Omega, 2^\Omega, P)$ is denoted $RV(\Omega, 2^\Omega, P)$ and it is used to represent the payoff of risky financial assets. Since the sample space $\Omega$ is finite, $RV(\Omega, 2^\Omega, P)$ is a finite dimensional vector space. We use $RV_+(\Omega, 2^\Omega, P)$ to represent of the cone of nonnegative random variables in $RV(\Omega, 2^\Omega, P)$. Introducing the inner product

$$\langle X, Y \rangle = \mathbb{E}[XY], \quad X, Y \in RV(\Omega, 2^\Omega, P),$$

$RV(\Omega, 2^\Omega, P)$ becomes a (finite dimensional) Hilbert space.

**Definition 2.1.** (Financial Market) *We say that $S_t = (S_t^0, S_t^1, \ldots, S_t^M), t = 0, 1$ is a financial market in a one period economy provided that $S_0 \in \mathbb{R}_+^{M+1}$ and $S_1 \in (0, \infty) \times RV_+(\Omega, 2^\Omega, P)^M$. Here $S_0^0 = 1, S_1^0 = R > 0$ represents a risk free bond with a positive return when $R > 1$. The rest of the components $S_t^m, m = 1, \ldots, M$ represent the price of the m-th risky financial asset at time t.*

We will use the notation $\widehat{S}_t = (S_t^1, \cdots, S_t^M)$ when we need to focus on the risky assets. We assume that $S_0$ is a constant vector representing the prices of the assets in this financial market at $t = 0$. The risk is modeled by assuming $\widehat{S}_1 = (S_1^1, \ldots, S_1^M)$ to be a nonnegative random vector on the probability space $(\Omega, 2^\Omega, P)$, that is $S_1^m \in RV_+(\Omega, 2^\Omega, P), m = 1, 2, \ldots, M$. A portfolio is a column vector $x \in \mathbb{R}^{M+1}$ whose components $x_m$ represent the share of the m-th asset in the portfolio and $S_t^m x_m$ is the portion of capital invested in asset $m$ at time $t$. Hence $x_0$ corresponds to the investment in the risk free bond and $\widehat{x} = (x_1, \ldots, x_M)^\top$ is the risky part.

We often need to restrict the selection of portfolios. For example, in many applications we consider only portfolios with unit initial cost, i.e. $S_0 \cdot x = 1$. Thus, the following definition.

**Definition 2.2.** (Admissible Portfolio) *We say that $A \subset \mathbb{R}^{M+1}$ is a set of* admissible portfolios *provided that $A$ is a nonempty closed and convex set. We say that $A$ is a set of* admissible portfolios with unit initial price *provided that $A$ is a closed convex subset of $\{x \in \mathbb{R}^{M+1} : S_0 \cdot x = 1\}$.*

## 2.2 Convex programming problems

Let $X$ be a finite dimensional Banach space. Recall that a set $C \subset X$ is convex if, for any $x, y \in C$ and $s \in [0, 1]$, $sx + (1 - s)y \in C$. For an extended valued function $f : X \to \mathbb{R} \cup \{+\infty\}$ we define its domain by

$$\text{dom}(f) := \{x \in X : f(x) < \infty\}$$

and its epigraph by

$$\text{epi}(f) := \{(x, r) \in X \times \mathbb{R} : r \geq f(x)\}.$$

We say $f$ is lower semicontinuous if $\text{epi}(f)$ is a closed set. The following proposition characterizes an epigraph of a function.

**Proposition 2.3.** (Characterization of Epigraph) *Let $F$ be a closed subset of $X \times \mathbb{R}$ such that $\inf\{r : (x, r) \in F\} > -\infty$ for all $x \in \mathbb{R}$. Then $F$ is the epigraph for a lower semicontinuous function $f : X \to (-\infty, \infty]$, i.e. $F = \text{epi}(f)$, if and only if*

$$(x, r) \in F \Rightarrow (x, r + k) \in F, \ \forall k > 0. \tag{2.1}$$

**Proof.** The key is to observe that, for a set $F$ with the structure in (2.1), a function

$$f(x) = \inf\{r : (x, r) \in F\} \tag{2.2}$$

is well defined and then $F = \text{epi}(f)$ holds. Q.E.D.

We say a function $f$ is convex if $\text{epi}(f)$ is a convex set. Alternatively, $f$ is convex if and only if, for any $x, y \in \text{dom}(f)$ and $s \in [0, 1]$,

$$f(sx + (1-s)y) \leq sf(x) + (1-s)f(y).$$

Consider $f : X \to [-\infty, +\infty)$. We say $f$ is concave when $-f$ is convex and we say $f$ is upper semicontinuous if $-f$ is lower semicontinuous. Define the hypograph of a function $f$ by

$$\text{hypo}(f) = \{(x, r) \in X \times \mathbb{R} : r \leq f(x)\}.$$

Then a symmetric version of Proposition 2.3 is

**Proposition 2.4.** (Characterization of Hypograph) *Let $F$ be a closed subset of $X \times \mathbb{R}$ such that $\sup\{r : (x, r) \in F\} < +\infty$ for all $x \in \mathbb{R}$. Then $F$ is the hypograph of an upper semicontinuous function $f : X \to [-\infty, \infty)$, i.e. $F = \text{hypo}(f)$, if and only if*

$$(x, r) \in F \Rightarrow (x, r - k) \in F, \ \forall k > 0. \tag{2.3}$$

*Moreover, the function $f$ can be defined by*

$$f(x) = \sup\{r : (x, r) \in F\}. \tag{2.4}$$

**Remark 2.5.** *The value of the function $f$ in Proposition 2.3 (Proposition 2.4) at a given point $x$ is $-\infty$ ($+\infty$) if and only if $\{x\} \times \mathbb{R} \subset F$.*

Since utility functions are concave and risk measures are usually convex, the analysis of a general trade-off between utility and risk naturally leads to a convex programming problem. The general form of such convex programming problems is

$$v(y, z) := \inf_{x \in X} [f(x) : g(x) \leq y, h(x) = z], \text{ for } y \in \mathbb{R}^M, z \in \mathbb{R}^N, \tag{2.5}$$

where $f$, $g$ and $h$ satisfy the following assumption.

**Assumption 2.6.** *Assume that $f : X \to \mathbb{R} \cup \{+\infty\}$ is a lower semicontinuous extended valued convex function, $g : X \to \mathbb{R}^M$ is a vector valued function with convex components, $\leq$ signifies componentwise minorization and $h : X \to \mathbb{R}^N$ is an affine mapping, for natural numbers $M, N$. Moreover, at least one of the components of $g$ has compact sublevel sets.*

Convex programming problems have nice properties due to the convex structure. We briefly recall the pertinent results related to convex programming. First the optimal value function $v$ is convex. This is a well-known result that can be found in standard books on convex analysis, e.g. [4]. It is, however, crucial for our applications below and, thus, we list it as a lemma and give a brief proof below for completeness.

**Proposition 2.7.** (Convexity of Optimal Value Function) *Let $f$, $g$ and $h$ satisfy Assumption 2.6. Then the optimal value function $v$ in the convex programming problem (2.5) is convex and lower semicontinuous.*

**Proof.** Consider $(y^i, z^i) \in \text{dom}(v), i = 1, 2$ in the domain of $v$ and an arbitrary $\varepsilon > 0$. We can find $x_\varepsilon^i$ feasible to the constraint of problem $v(y^i, z^i)$ such that

$$f(x_\varepsilon^i) < v(y^i, z^i) + \varepsilon, \ i = 1, 2. \tag{2.6}$$

Now for any $\lambda \in [0, 1]$, we have

$$\begin{aligned} f(\lambda x_\varepsilon^1 + (1-\lambda) x_\varepsilon^2) &\leq \lambda f(x_\varepsilon^1) + (1-\lambda) f(x_\varepsilon^2) \\ &< \lambda v(y^1, z^1) + (1-\lambda) v(y^2, z^2) + \varepsilon. \end{aligned} \tag{2.7}$$

It is easy to check that $\lambda x_\varepsilon^1 + (1-\lambda) x_\varepsilon^2$ is feasible for the problem $v(\lambda(y^1, z^1) + (1-\lambda)(y^2, z^2))$. Thus, $v(\lambda(y^1, z^1) + (1-\lambda)(y^2, z^2)) \leq f(\lambda x_\varepsilon^1 + (1-\lambda) x_\varepsilon^2)$. Combining with inequality (2.7) and letting $\varepsilon \to 0$ we arrive at

$$v(\lambda(y^1, z^1) + (1-\lambda)(y^2, z^2)) \leq \lambda v(y^1, z^1) + (1-\lambda) v(y^2, z^2),$$

that is to say $v$ is convex.

The lower semicontinuity of $v$ is easier to verify. Q.E.D.

By and large, there are two (equivalent) general approaches to help solving a convex programming problem: by using the related dual problem and by using Lagrange multipliers. The two methods are equivalent in the sense that a solution to the dual problem is exactly a Lagrange multiplier (see [5]). Using Lagrange multipliers is more accessible to practitioners outside the special area of convex analysis. We will take this approach. The Lagrange multipliers method tells us that under mild assumptions we can expect there exists a Lagrange multiplier $\lambda = (\lambda_y, \lambda_z)$ with $\lambda_y \geq 0$ such that $\bar{x}$ is a solution to the convex programming problem (2.5) if and only if it is a solution to the unconstrained problem of minimizing

$$L(x, \lambda) := f(x) + \langle \lambda, (g(x) - y, h(x) - z) \rangle = f(x) + \langle \lambda_y, g(x) - y \rangle + \langle \lambda_z, h(x) - z \rangle. \tag{2.8}$$

The function $L(x, \lambda)$ is called the Lagrangian. To understand why and when does a Lagrange multiplier exist, we need to recall the definition of the subdifferential.

**Definition 2.8.** (Subdifferential) *Let $X$ be a finite dimensional Banach space and $X^*$ its dual space. The subdifferential of a lower semicontinuous convex function $\phi : X \to \mathbb{R} \cup \{+\infty\}$ at $x \in \text{dom}(\phi)$ is defined by*

$$\partial \phi(x) = \{x^* \in X^* : \phi(y) - \phi(x) \geq \langle x^*, y - x \rangle \ \forall y \in X\}.$$

Geometrically, an element of the subdifferential gives us the normal vector of a support hyperplane for the convex function at the relevant point. It turns out that Lagrange multipliers of problem (2.5) are simply the negative of elements of the subdifferential of $v$. We summarize and prove the sufficiency in the lemma below which we will actually use.

**Theorem 2.9.** (Lagrange Multiplier) *Let $v : \mathbb{R}^M \times \mathbb{R}^N \to \mathbb{R} \cup \{+\infty\}$ be the optimal value function of the constrained optimization problem (2.5) with $f, g$ and $h$ satisfying Assumption 2.6. Suppose that, for fixed $(y, z) \in \mathbb{R}^M \times \mathbb{R}^N$, $-\lambda = -(\lambda_y, \lambda_z) \in \partial v(y, z)$ and $\bar{x}$ is a solution of (2.5). Then*

(i) $\lambda_y \geq 0$,

(ii) *the Lagrangian $L(x, \lambda)$ defined in (2.8) attains a global minimum at $\bar{x}$, and*

(iii) $\lambda$ *satisfies the complementary slackness condition*

$$\langle \lambda, (g(\bar{x}) - y, h(\bar{x}) - z) \rangle = \langle \lambda_y, g(\bar{x}) - y \rangle = 0. \tag{2.9}$$

**Proof.** Observe that $v(y, z)$ is a nonincreasing function with respect to the minorization $\leq$ in $y$. Using $-\lambda \in \partial v(y, z)$, for any vector $\Delta y \geq 0$, we have

$$0 \geq v(y + \Delta y, z) - v(y, z) \geq \langle -\lambda, (\Delta y, 0) \rangle.$$

It follows that $\lambda_y \geq 0$ verifying (i).

By the definition of the subdifferential and the fact that $v(g(\bar{x}), h(\bar{x})) = v(y, z)$, we then have

$$0 = v(g(\bar{x}), h(\bar{x})) - v(y, z) \geq \langle -\lambda, (g(\bar{x}) - y, h(\bar{x}) - z) \rangle \geq 0.$$

It follows that the complementary slackness condition

$$\langle \lambda, (g(\bar{x}) - y, h(\bar{x}) - z) \rangle = 0 \tag{2.10}$$

in (iii) holds.

Finally, by the definition of the subdifferential we have

$$v(g(x), h(x)) - v(y, z) \geq \langle -\lambda, (g(x) - y, h(x) - z) \rangle.$$

Thus, for any $x$,

$$\begin{aligned} L(x, \lambda) &= f(x) + \langle \lambda, (g(x) - y, h(x) - z) \rangle \\ &\geq v(g(x), h(x)) + \langle \lambda, (g(x) - y, h(x) - z) \rangle \\ &\geq v(y, z) \end{aligned} \tag{2.11}$$

Using the fact that $\bar{x}$ is a solution to problem in (2.5) and the complementary slackness condition (2.10) we have

$$v(y, z) = f(\bar{x}) = f(\bar{x}) + \langle \lambda, (g(\bar{x}) - y, h(\bar{x}) - z) \rangle = L(\bar{x}, \lambda). \tag{2.12}$$

Combining (2.11) and (2.12) verifies (ii). Q.E.D.

**Remark 2.10.** By Theorem 2.9 Lagrange multipliers exist when (2.5) has a solution $\bar{x}$ and $\partial v(y,z) \neq \emptyset$. Calculating $\partial v(y,z)$ requires to know the value of $v$ in a neighborhood of $(y,z)$ and is not realistic. Fortunately, the well-known Fenchel-Rockafellar theorem (see e.g. [4]) tells us when $(y,z)$ belongs to the relative interior of $\text{dom}(v)$, then $\partial v(y,z) \neq \emptyset$. This is a very useful sufficient condition. A particularly useful special case is the Slater condition (see also [4]): when there is only an inequality constraint $g(x) \leq y$, if there exists $x \in \text{dom}(f)$ such that $g(x) < y$ implies already that $\partial v(y) \neq \emptyset$.

# 3 Efficient trade-off between risk and utility

We consider the financial market described in Definition 2.1 and consider a set of admissible portfolios $A \subset \mathbb{R}^{M+1}$ (see Definition 2.2). The payoff of each portfolio $x \in A$ at time $t=1$ is $S_1 \cdot x$. The merit of a portfolio $x$ is often judged by its expected utility $\mathbb{E}[u(S_1 \cdot x)]$ where $u$ is an increasing concave utility function. The increasing property of $u$ models the more payoff the better. The concavity reflects the fact that with the increase of payoff, its marginal utility to an investor decreases. On the other hand investors are often sensitive to the risk of a portfolio which can be gauged by a risk measure. Because diversification reduces risk, the risk measure should be a convex function.

## 3.1 Technical Assumptions

Some standard assumptions on the utility and risk functions are often needed in the more technical discussion below. We collect them here.

**Assumption 3.1.** (Conditions on Risk Measure) *Consider a continuous risk function* $\mathfrak{r}: A \to [0,+\infty)$ *where $A$ is a set of admissible portfolios according to Definition 2.2. We will often refer to some of the following assumptions.*

(r1) (Riskless Asset Contributes No risk) *The risk measure $\mathfrak{r}(x) = \widehat{\mathfrak{r}}(\widehat{x})$ is a function of only the risky part of the portfolio, where $x = (x_0, \widehat{x})^\top$.*

(r1n) (Normalization) *There is at least one portfolio of purely bonds in $A$. Furthermore, $\mathfrak{r}(x) = 0$ if and only if $x$ contains only riskless bonds, i.e. $x = (x_0, \widehat{0})^\top$ for some $x_0 \in \mathbb{R}$.*

(r2) (Diversification Reduces Risk) *The risk function $\mathfrak{r}$ is convex.*

(r2s) (Diversification Strictly Reduces Risk) *The risk function $\widehat{\mathfrak{r}}$ is strictly convex.*

(r3) (Positive homogeneous) *For $t > 0$, $\widehat{\mathfrak{r}}(t\widehat{x}) = t\widehat{\mathfrak{r}}(\widehat{x})$.*

**Remark 3.2.** (Deviation measure) *A risk measure satisfying assumptions (r1), (r1n), (r2) and (r3) is strongly related to a* deviation measure *in [20]. It is also related to the coherent risk measure introduced in [1].*

**Assumption 3.3.** (Conditions on Utility Function) *Utility functions $u : \mathbb{R} \to \mathbb{R} \cup \{-\infty\}$ are usually assumed to satisfy some of the following properties.*

(u1) (Profit Seeking) *The utility function $u$ is an increasing function.*

(u2) (Diminishing Marginal Utility) *The utility function $u$ is concave.*

(u2s) (Strict Diminishing Marginal Utility) *The utility function $u$ is strictly concave.*

(u3) (Bankrupcy Forbidden) *For $t < 0$, $u(t) = -\infty$.*

(u4) (Unlimited Growth) *For $t \to +\infty$, we have $u(t) \to +\infty$.*

Another important condition which often appears in the financial literature is no arbitrage.

**Definition 3.4.** (No Arbitrage) *We say a portfolio $x \in \mathbb{R}^{M+1}$ is an* arbitrage *on the financial market $S$ if*

$$(S_1 - RS_0) \cdot x \geq 0 \text{ and } (S_1 - RS_0) \cdot x \neq 0.$$

*We say market $S_t$ has* no arbitrage *if there does not exist any arbitrage portfolio for the financial market $S_t$.*

An arbitrage is a way to make return above the risk free rate without taking any risk of losing money. If such an opportunity exists then investors will try to take advantage of it. In this process they will bid up the price of the risky assets and cause the arbitrage opportunity to disappear. For this reason, usually people assume a financial market does not contain any arbitrage.

The following is a weaker requirement than arbitrage:

**Definition 3.5.** (No Nontrivial Riskless Portfolio) *We say a portfolio $x \in \mathbb{R}^{M+1}$ is riskless if*

$$(S_1 - RS_0) \cdot x \geq 0.$$

*We say the market has* no nontrivial riskless portfolio *if there does not exist a riskless portfolio $x$ with $\widehat{x} \neq \widehat{0}$.*

A trivial riskless portfolio of investing everything in the riskless asset $S_t^0$ always exists. A nontrivial riskless portfolio, however, is not to be expected and we will often use this assumption.

It turns out that the difference between no nontrivial riskless portfolio and no arbitrage is exactly the following:

**Definition 3.6.** (Nontrivial Bond Replicating Portfolio) *We say that $x = (x_0, \widehat{x})^\top$ is a nontrivial bond replicating portfolio if $\widehat{x} \neq \widehat{0}$ and*

$$(S_1 - RS_0) \cdot x = 0.$$

The three conditions in Definitions 3.4, 3.5 and 3.6 are related as follows:

**Proposition 3.7.** *Consider financial market $S_t$ of Definition 2.1. There is no nontrivial riskless portfolio in $S_t$ if and only if $S_t$ has no arbitrage portfolio and no nontrivial bond replicating portfolio.*

**Proof.** The conclusion follows directly from Definitions 3.4, 3.5 and 3.6.    Q.E.D.

**Corollary 3.8.** *No nontrivial riskless portfolio implies no arbitrage portfolio.*

Assuming the financial market has no arbitrage then no nontrivial riskless portfolio is equivalent to no nontrivial bond replicating portfolio and has the following characterization.

**Theorem 3.9.** (Characterization of no Nontrivial Bond Replicating Portfolio) *Assuming the financial market $S_t$ in Definition 2.1 has no arbitrage. Then the following assertions are equivalent:*

(i) *There is no nontrivial bond replicating portfolio.*

(ii) *For every nontrivial portfolio $x$ with $\widehat{x} \neq \widehat{0}$, there exists some $\omega \in \Omega$ such that*

$$(S_1(\omega) - RS_0) \cdot x < 0. \tag{3.1}$$

(ii*) *For every risky portfolio $\widehat{x} \neq \widehat{0}$, there exists some $\omega \in \Omega$ such that*

$$(\widehat{S}_1(\omega) - R\widehat{S}_0) \cdot \widehat{x} < 0. \tag{3.2}$$

(iii) *The matrix*

$$G := \begin{bmatrix} S_1^1(\omega_1) - RS_0^1 & S_1^2(\omega_1) - RS_0^2 & \ldots & S_1^M(\omega_1) - RS_0^M \\ S_1^1(\omega_2) - RS_0^1 & S_1^2(\omega_2) - RS_0^2 & \ldots & S_1^M(\omega_2) - RS_0^M \\ \vdots & \vdots & \vdots & \vdots \\ S_1^1(\omega_N) - RS_0^1 & S_1^2(\omega_N) - RS_0^2 & \ldots & S_1^M(\omega_N) - RS_0^M \end{bmatrix} \in \mathbb{R}^{N \times M} \tag{3.3}$$

*has rank $M$, in particular $N \geq M$.*

**Proof.** We use a cyclic proof. (i)→ (ii): If (ii) fails then $(S_1 - RS_0) \cdot x \geq 0$ for some nontrivial $x$. By (i) $x$ must be an arbitrage, which is a contradiction. (ii)→ (ii*): obvious. (ii*)→ (iii): If (iii) is not true then $G \cdot \widehat{x} = 0$ has a nontrivial solution which is a contradiction to (3.2). (iii)→ (i): Assume that there exists a portfolio $x^*$ with $\widehat{x}^* \neq \widehat{0}$ which replicates the bond. Then $(S_1 - RS_0) \cdot x^* = 0$. This implies that $(\widehat{S}_1 - R\widehat{S}_0) \cdot \widehat{x}^* = 0$ so that $G\widehat{x}^* = 0$ which contradicts (iii).    Q.E.D.

A rather useful corollary of Theorem 3.9 is that any of the conditions (i)–(iii) of that theorem ensures the covariance matrix of the risky assets to be positive definite.

**Corollary 3.10.** (Positive Definite Covariance Matrix) *Assume the financial market $S_t$ in Definition 2.1 has no nontrivial riskless portfolio. Then the covariant matrix of the risky assets*

$$\begin{aligned}\Sigma &:= \mathbb{E}[(\widehat{S}_1 - \mathbb{E}(\widehat{S}_1))^\top (\widehat{S}_1 - \mathbb{E}(\widehat{S}_1))] \\ &= (\mathbb{E}[(S_1^i - \mathbb{E}(S_1^i))(S_1^j - \mathbb{E}(S_1^j))])_{i,j=1,\dots,M},\end{aligned} \quad (3.4)$$

*is positive definite.*

**Proof.** We note that under the assumption of the corollary, for any nontrivial risky portfolio $\widehat{x}$, $\widehat{S}_1 \cdot \widehat{x}$ cannot be a constant. Otherwise, $(\widehat{S}_1 - R\widehat{S}_0) \cdot \widehat{x}$ would be a constant which contradicts $S_t$ has no nontrivial riskless portfolio. It follows that for any nontrivial risky portfolio $\widehat{x}$,

$$Var(\widehat{S}_1 \cdot \widehat{x}) = \widehat{x}^\top \Sigma \widehat{x} > 0.$$

Thus, $\Sigma$ is positive definite. Q.E.D.

**Remark 3.11.** *Corollary 3.10 shows that the standard deviation as a risk measure satisfies the properties (r1), (r1n), (r2) and (r3) in Assumption 3.1.*

## 3.2 Efficient Frontier for the Risk-Utility trade-off

We note that to increase the utility one often has to take on more risk and as a result the risk increases. The converse is also true. For example, if one allocates all the capital to the riskless bond then there will be no risk but the price to pay is that one has to forgo all the opportunities to get a high payoff on risky assets so as to reduce the expected utility. Thus, the investment decision of selecting an appropriate portfolio becomes one of trading-off between the portfolio's expected return and risk. To understand such a trade-off we define, for a set of admissible portfolios $A \subset \mathbb{R}^{M+1}$ in Definition 2.2, the set

$$\mathcal{G}(\mathfrak{r}, u; A) := \{(r, \mu) : \exists x \in A \text{ s.t. } r \geq \mathfrak{r}(x), \mu \leq \mathbb{E}[u(S_1 \cdot x)]\} \subset \mathbb{R}^2, \quad (3.5)$$

on the two dimensional risk-expected utility space for a given risk measure $\mathfrak{r}$ and utility $u$. Given a financial market $S_t$ and a portfolio $x$, we often measure risk by observing $S_1 \cdot x$.

**Corollary 3.12.** (Induced Risk Measure) *(a) Fixing a financial market $S_t$ as in Definition 2.1. Suppose that $\rho : RV(\Omega, 2^\Omega, P) \to [0, +\infty)$ is a lower semicontinuous, convex and positive homogeneous function. Moreover, assume that $\rho(S_1 \cdot x) = \rho(\widehat{S}_1 \cdot \widehat{x})$. Then $\mathfrak{r} : A \to [0, +\infty)$, $\mathfrak{r}(x) := \rho(S_1 \cdot x)$ is a lower semicontinuous risk measure satisfying properties* (r1), (r2) *and* (r3) *in Assumption 3.1.*

*(b) If the financial market $S_t$ has no nontrivial riskless portfolio and $\rho$ is strictly convex then for a set $A$ of admissible portfolios with unit initial cost, $\widehat{\mathfrak{r}} : A \to [0, +\infty)$ satisfies* (r2s) *in Assumption 3.1.*

**Proof.** Since $x \to S_1 \cdot x$ is a linear mapping, the risk measure $\mathfrak{r}$ inherits the properties of $\rho$ so that it satisfies properties (r1), (r2) and (r3) in Assumption 3.1. One sufficient condition for $\hat{\mathfrak{r}}$ to preserve the strict convexity of $\rho$ is that the matrix $G$ in (3.3) is of full rank since all portfolios have unit initial cost. It follows from Theorem 3.9 that this condition follows from no nontrivial riskless portfolio in the financial market $S_t$. Q.E.D.

**Remark 3.13.** *The following are two sufficient conditions ensuring $\rho(S_1 \cdot x) = \rho(\widehat{S}_1 \cdot \widehat{x})$ that are easy to verify:*

(1) *When $\rho$ is invariant under adding constants, i.e., $\rho(X) = \rho(X + c)$, for any $X \in RV(\Omega, 2^\Omega, P)$ and $c \in \mathbb{R}$. A useful example is when $\rho$ is the standard deviation.*

(2) *When $\rho$ is restricted to a set of admissible portfolios $A$ with unit initial cost. In this case we can see that*

$$\widehat{\mathfrak{r}}(\widehat{x}) := \rho(R + (\widehat{S}_1 - R\widehat{S}_0) \cdot \widehat{x}) = \rho(S_1 \cdot x). \tag{3.6}$$

Similarly, we are interested in when the expected utility $x \mapsto \mathbb{E}[u(S_1 \cdot x)]$ of $S_1 \cdot x$ is strictly concave in $x$. Below is a set of useful sufficient conditions.

**Lemma 3.14.** *(Strict Concavity of Expected Utility) Assume that*

(a) *the financial market $S_t$ has no nontrivial riskless portfolio,*

(b) *the utility function $u$ satisfies condition (u2s) in Assumption 3.3, and*

(c) *$A$ is a set of admissible portfolios with unit initial cost as in Definition 2.2.*

*Then the expected utility $\mathbb{E}[u(S_1 \cdot x)]$ as a function of the portfolio $x$ is strictly concave on $A$.*

**Proof.** Since $u$ is concave so is $x \mapsto \mathbb{E}[u(S_1 \cdot x)]$. To prove that this function is strictly concave on $A$, consider two distinct portfolios $x_1, x_2 \in A$. By assumption (c), both $x_1$ and $x_2$ have unit initial cost and thus $\widehat{x}_1 \neq \widehat{x}_2$. Assumption (a) and Proposition 3.7 implies that for the matrix $G$ defined in (3.3), $G\widehat{x}_1 \neq G\widehat{x}_2$. Thus, using again the fact that both $x_1$ and $x_2$ have unit initial cost, we have

$$S_1 \cdot x_1 = R + (\widehat{S}_1 - R\widehat{S}_0) \cdot \widehat{x}_1 \neq R + (\widehat{S}_1 - R\widehat{S}_0) \cdot \widehat{x}_2 = S_1 \cdot x_2.$$

The strictly concavity of $x \to \mathbb{E}[u(S_1 \cdot x)]$ now follows from the strict concavity of the utility function $u$ as assumed in (b). Q.E.D.

When $\mathfrak{r}(x) = \rho(S_1 \cdot x)$ is induced by $\rho$ as in Corollary 3.12 we also use the notation $\mathcal{G}(\rho, u, A)$. Clearly, if $A' \subset A$ then $\mathcal{G}(\mathfrak{r}, u; A') \subset \mathcal{G}(\mathfrak{r}, u; A)$. The following assumption will be needed in concrete applications.

**Assumption 3.15.** *(Compact Level Sets) Either (a) for each $\mu \in \mathbb{R}$, $\{x \in \mathbb{R}^{M+1} : \mu \leq \mathbb{E}[u(S_1 \cdot x)], x \in A\}$ is compact or (b) for each $r \in \mathbb{R}$, $\{x \in \mathbb{R}^{M+1} : r \geq \mathfrak{r}(x), x \in A\}$ is compact.*

**Proposition 3.16.** *Assume that $A$ is a set of admissible portfolios as in Definition 2.2. We claim: (a) Assume that the risk measure $\mathfrak{r}$ satisfies (r2) in Assumption 3.1 and the utility function $u$ satisfies (u2) in Assumption 3.3. Then set $\mathcal{G}(\mathfrak{r}, u; A)$ is convex and $(r, \mu) \in \mathcal{G}(\mathfrak{r}, u; A)$ implies that, for any $k > 0$, $(r + k, \mu) \in \mathcal{G}(\mathfrak{r}, u; A)$ and $(r, \mu - k) \in \mathcal{G}(\mathfrak{r}, u; A)$. (b) Assume furthermore that Assumption 3.15 holds. Then $\mathcal{G}(\mathfrak{r}, u; A)$ is closed.*

**Proof.** (a) The property $(r, \mu) \in \mathcal{G}(\mathfrak{r}, u; A)$ implies that, for any $k > 0$, $(r + k, \mu) \in \mathcal{G}(\mathfrak{r}, u; A)$ and $(r, \mu - k) \in \mathcal{G}(\mathfrak{r}, u; A)$ follows directly from the definition of $\mathcal{G}(\mathfrak{r}, u; A)$.

Suppose that $(r_1, \mu_1), (r_2, \mu_2) \in \mathcal{G}(\mathfrak{r}, u; A)$ and $s \in [0, 1]$. Then there exists $x^1, x^2 \in A$ such that
$$r_i \geq \mathfrak{r}(x^i) \text{ and } \mu_i \leq \mathbb{E}[u(S_1 \cdot x^i)], i = 1, 2.$$
Then convexity of $\mathfrak{r}$ in $x$ yields
$$sr_1 + (1-s)r_2 \geq s\mathfrak{r}(x^1) + (1-s)\mathfrak{r}(x^2) \geq \mathfrak{r}(sx^1 + (1-s)x^2),$$
and (u2) gives
$$s\mu_1 + (1-s)\mu_2 \leq s\mathbb{E}[u(S_1 \cdot x^1)] + (1-s)\mathbb{E}[u(S_1 \cdot x^2)] \leq \mathbb{E}[u(S_1 \cdot (sx^1 + (1-s)x^2))].$$
Thus,
$$s(r_1, \mu_1) + (1-s)(r_2, \mu_2) \in \mathcal{G}(\mathfrak{r}, u; A)$$
so that $\mathcal{G}(\mathfrak{r}, u; A)$ is convex.

(b) Suppose that $(r_n, \mu_n) \to (r, \mu)$, for a sequence in $\mathcal{G}(\mathfrak{r}, u; A)$. Then there exists a sequence $x^n \in A$ such that
$$r_n \geq \mathfrak{r}(x^n) \text{ and } \mu_n \leq \mathbb{E}[u(S_1 \cdot x^n)]. \tag{3.7}$$
By Assumption 3.15 a subsequence of $x^n$ (denoted again by $x^n$) converges to, say, $\bar{x} \in A$. Taking limits in (3.7) we arrive at
$$r \geq \mathfrak{r}(\bar{x}) \text{ and } \mu \leq \mathbb{E}[u(S_1 \cdot \bar{x})]. \tag{3.8}$$
Thus, $(r, \mu) \in \mathcal{G}(\mathfrak{r}, u; A)$ and hence $\mathcal{G}(\mathfrak{r}, u; A)$ is a closed set. Q.E.D.

Now we can represent a portfolio $x \in A \subset \mathbb{R}^{M+1}$ as a point $(\mathfrak{r}(x), \mathbb{E}[u(S_1 \cdot x)]) \in \mathcal{G}(\mathfrak{r}, u; A)$ in the two dimensional risk-expected utility space. Investors prefer portfolios with lower risk if the expected utility is the same or with higher expected utility given the same level of risk.

**Definition 3.17.** (*Efficient Portfolio*) *We say that a portfolio $x \in A$ is* Pareto efficient *provided that there does not exist any portfolio $x' \in A$ such that either*
$$\mathfrak{r}(x') \leq \mathfrak{r}(x) \text{ and } \mathbb{E}[u(S_1 \cdot x')] > \mathbb{E}[u(S_1 \cdot x)]$$
*or*
$$\mathfrak{r}(x') < \mathfrak{r}(x) \text{ and } \mathbb{E}[u(S_1 \cdot x')] \geq \mathbb{E}[u(S_1 \cdot x)].$$

**Definition 3.18.** (Efficient Frontier) *We call the set of images of all efficient portfolios in the two dimensional risk-expected utility space the* efficient frontier *and denote it by* $\mathcal{G}_{eff}(\mathfrak{r}, u; A)$.

The next theorem characterizes efficient portfolios in the risk-expected utility space.

**Theorem 3.19.** (Efficient Frontier) *Efficient portfolios represented in the two dimensional risk-expected utility space are all located in the (non vertical or horizontal) boundary of the set* $\mathcal{G}(\mathfrak{r}, u; A)$.

**Proof.** If a portfolio $x$ represented in the risk-expected utility space as $(r, \mu)$ is not on the (non vertical or horizontal) boundary of the $\mathcal{G}(\mathfrak{r}, u; A)$, then for $\varepsilon$ small enough we have either $(r - \varepsilon, \mu) \in \mathcal{G}(\mathfrak{r}, u; A)$ or $(r, \mu + \varepsilon) \in \mathcal{G}(\mathfrak{r}, u; A)$. This means $x$ can be improved. Q.E.D.

The following relationship is straightforward but very useful.

**Theorem 3.20.** (Efficient Frontier of Subsystem) *Consider admissible portfolios $A, B$. If $B \subset A$ then $\mathcal{G}_{eff}(\mathfrak{r}, u; A) \cap \mathcal{G}(\mathfrak{r}, u; B) \subset \mathcal{G}_{eff}(\mathfrak{r}, u; B)$.*

**Proof.** The conclusion directly follows from $\mathcal{G}(\mathfrak{r}, u; B) \subset \mathcal{G}(\mathfrak{r}, u; A)$. Q.E.D.

**Remark 3.21.** (Empty Efficient Frontier) *If $(\alpha, \widehat{0}) \in A$ for all $\alpha \in \mathbb{R}$ and the increasing utility function $u$ has no upper bound then for any risk measure $\mathfrak{r}$ satisfying (r1) and (r1n) in Assumption 3.1, $\{0\} \times \mathbb{R} \subset \mathcal{G}(\mathfrak{r}, u; A)$. By Proposition 3.16 $[0, +\infty) \times \mathbb{R} \subset \mathcal{G}(\mathfrak{r}, u; A)$ which implies that $\mathcal{G}_{eff}(\mathfrak{r}, u; A) = \emptyset$. Thus, practically meaningful $\mathcal{G}(\mathfrak{r}, u; A)$ always correspond to sets of admissible portfolios $A$ such that the initial cost $S_0 \cdot x$ for all $x \in A$ is limited. Moreover, if the initial cost has a range and riskless bonds are included in the portfolio, then we will see a vertical line segment on the $\mu$ axis and the efficient portfolio corresponds to the upper bound of this vertical line segments. Thus, it suffices to consider sets of portfolios $A$ with unit initial cost.*

## 3.3 Representation of Efficient Frontier

In view of Remark 3.21, in this section we will consider a set of admissible portfolios $A$ with unit initial cost as in Definition 2.2. By Proposition 3.16 we can view the set $\mathcal{G}(\mathfrak{r}, u; A)$ as an epigraph on the expected utility-risk space or a hypograph on the risk-expected utility space. By Propositions 2.3 and 2.4, the set $\mathcal{G}(\mathfrak{r}, u; A)$ naturally defines two functions

$$\begin{aligned}\gamma(\mu) &:= \inf\{r : (r, \mu) \in \mathcal{G}(\mathfrak{r}, u; A)\} \\ &= \inf\{\mathfrak{r}(x) : \mathbb{E}[u(S_1 \cdot x)] \geq \mu, x \in A\},\end{aligned} \quad (3.9)$$

and

$$\begin{aligned}\nu(r) &:= \sup\{\mu : (r, \mu) \in \mathcal{G}(\mathfrak{r}, u; A)\} \\ &= \sup\{\mathbb{E}[u(S_1 \cdot x)] : \mathfrak{r}(x) \leq r, x \in A\}.\end{aligned} \quad (3.10)$$

**Proposition 3.22.** (Function Related to the Efficient Frontier) *Assume that, the risk measure $\mathfrak{r}$ satisfies (r2) in Assumption 3.1 and the utility function u satisfies (u2) in Assumption 3.3. Furthermore, assume that Assumption 3.15 holds for a set of admissible portfolios A with unit initial cost. Then the functions $\mu \mapsto \gamma(\mu)$ and $r \mapsto \nu(r)$ are increasing lower semicontinuous convex and increasing upper semicontinuous concave, respectively.*

**Proof.** The conclusion follows directly from Propositions 2.3 and 2.4 since $\mathcal{G}(\mathfrak{r}, u, A)$ is closed and convex according to Proposition 3.16.

Alternatively, we can also directly apply Proposition 2.7 to the second representation in (3.9) and (3.10) to derive the convexity and concavity of $\gamma$ and $\nu$, respectively.

The increasing property of $\gamma$ and $\nu$ follows directly from the second representation in (3.9) and (3.10), respectively. Q.E.D.

It also follows

**Corollary 3.23.** (Representation of Efficient Frontier) *Assume that the risk measure $\mathfrak{r}$ satisfies condition (r2) in Assumption 3.1 and the utility function u satisfies condition (u2) in Assumption 3.3. Then, for any set of admissible portfolios A with unit initial cost as defined in Definition 2.2, Pareto efficient portfolios $\mathcal{G}_{eff}(\mathfrak{r}, u; A)$ represented in the expected utility-risk space are all located on the graph of $\gamma$ or $\nu$, i.e.,*

$$\mathcal{G}_{eff}(\mathfrak{r}, u; A) = \operatorname{graph} \gamma(\mu) = \operatorname{graph} \nu(r).$$

## 3.4 Efficient Portfolios

We have seen that the efficient trade-off between risk and expected utility of a portfolio can be represented as the graph of a lower semicontinuous convex function $\mu \mapsto \gamma(\mu)$ that relates the level of expected return $\mu$ to a minimum risk. Alternatively, these points in the expected utility-risk space can also be represented as the graph of an upper semicontinuous concave function $r \mapsto \nu(r)$ that relates the level of risk $r$ to a maximum possible utility. We now turn to analyze how the corresponding efficient portfolios behave. Ideally we would want that each point on the efficient trade-off frontier corresponds to exactly one portfolio. For this purpose we need additional assumptions on risk measures and utility functions.

**Theorem 3.24.** (Efficient Portfolio Path) *Assume that the financial market $S_t$ defined in Definition 2.1 has no nontrivial riskless portfolio and that A is a set of admissible portfolios with unit initial cost as in Definition 2.2. We also assume Assumption 3.15 holds. In addition, suppose that one of the following conditions holds:*

(c1) *The risk measure $\mathfrak{r}$ satisfies conditions (r1) and (r2s) in Assumption 3.1 and the utility function satisfies conditions (u1) and (u2) in Assumption 3.3.*

(c2) *The risk measure $\mathfrak{r}$ satisfies conditions (r1) and (r2) in Assumption 3.1 and the utility function satisfies conditions (u1) and (u2s) in Assumption 3.3.*

Then, in case there exists some $x \in A$ with $\mathbb{E}[u(S_1 \cdot x)]$ finite, we can define

$$\mu_{\max} := \sup\{\mathbb{E}[u(S_1 \cdot x)], x \in A\} > -\infty, \tag{3.11}$$

$$r_{\min} := \inf\{\mathfrak{r}(x), x \in A\} \in [0, +\infty), \tag{3.12}$$

$$\mu_{\min} := \lim_{r \downarrow r_{\min}} \sup\{\mathbb{E}[u(S_1 \cdot x)] : \mathfrak{r}(x) \leq r, x \in A\}, \tag{3.13}$$

and

$$r_{\max} := \lim_{\mu \uparrow \mu_{\max}} \inf\{\mathfrak{r}(x) : \mathbb{E}[u(S_1 \cdot x)] \geq \mu, x \in A\} \tag{3.14}$$

and claim the following:

(a) *For $\mu \in (\mu_{\min}, \mu_{\max})$ there exists exactly one portfolio $x(\mu)$ on the efficient frontier $\mathcal{G}_{eff}(\mathfrak{r}, u, A)$ which corresponds to $(\gamma(\mu), \mu)$. Moreover, the mapping $\mu \to x(\mu)$ is continuous on $(\mu_{\min}, \mu_{\max})$. Furthermore, when $\mu_{\max}$ and/or $\mu_{\min}$ are/is attained by some $x \in A$ the above statement holds on the interval $(\mu_{\min}, \mu_{\max}]$, $[\mu_{\min}, \mu_{\max})$ or $[\mu_{\min}, \mu_{\max}]$ .*

(b) *For $r \in (r_{\min}, r_{\max})$ there exists exactly one portfolio $y(r)$ on the efficient frontier $\mathcal{G}_{eff}(\mathfrak{r}, u, A)$ which corresponds to $(r, \nu(r))$. Moreover, the mapping $r \to y(r)$ is continuous on $(r_{\min}, r_{\max})$. Furthermore, when $r_{\min}$ is a minimum and/or $r_{\max}$ are/is attained by some $x \in A$, the above statement holds on the interval $[r_{\min}, r_{\max})$, $(r_{\min}, r_{\max}]$, or $[r_{\min}, r_{\max}]$.*

(c) *If in addition, $\mathfrak{r}$ satisfies (r1n) in Assumption 3.1 then $r_{\min} = 0$, $\mu_{\min} = u(R)$ and $x(\mu_{\min}) = y(r_{\min}) = (1, \widehat{0})^\top$ (see Figure 2).*

**Proof.** (a) We focus on the case when condition (c1) is satisfied and will comment on the modifications needed for the similar case when (c2) is satisfied.

Consider $\mu \in (\mu_{\min}, \mu_{\max})$. Then we can find a portfolio $\bar{x} \in A$ with $\mathbb{E}[u(S_1 \cdot \bar{x})] \geq \mu$. By (3.12) $r_{\min} \leq \mathfrak{r}(\bar{x})$. Thus, the set $A_\mu := \{x : \mu \leq \mathbb{E}[u(S_1 \cdot x)], \mathfrak{r}(x) \leq \mathfrak{r}(\bar{x}), x \in A\}$ is nonempty. Moreover, Assumption 3.15 ensures that $A_\mu$ is compact. It follows that there exists at least one portfolio $x(\mu)$ such that

$$\mathfrak{r}(x(\mu)) = \inf\{\mathfrak{r}(x) : x \in A_\mu\} = \inf\{\mathfrak{r}(x) : \mu \leq \mathbb{E}[u(S_1 \cdot x)], x \in A\}.$$

Clearly, $x(\mu)$ corresponds to the point $(\gamma(\mu), \mu)$ on the efficient frontier $\mathcal{G}_{eff}(\mathfrak{r}, u, A)$.

Next we show the portfolio $x(\mu)$ is unique. Suppose that portfolios $x^1 \neq x^2$ both correspond to $(\gamma(\mu), \mu)$ and belong to $A$. Then we must have $\mathfrak{r}(x^1) = \mathfrak{r}(x^2) = \gamma(\mu)$ and $\mathbb{E}[u(S_1 \cdot x^i)] \geq \mu, x^i \in A, i = 1, 2$. Since $A$ is convex, $x^* = (x^1 + x^2)/2 \in A$. Conditions (r2s) and (u2) imply that $\mathbb{E}[u(S_1 \cdot x^*)] \geq \mu$ and due to the strict convexity of $\widehat{\mathfrak{r}}$ and (r1), $\mathfrak{r}(x^*) = \widehat{\mathfrak{r}}(\widehat{x}^*) < \gamma(\mu)$, a contradiction. Thus, the mapping $\mu \to x(\mu)$ is well defined.

Finally, we show the continuity of $x(\mu)$ by contradiction. Suppose this mapping is discontinuous at $\mu_0$. Then, for a fixed positive number $\varepsilon_0 > 0$, there exists a sequence $\mu_n \to \mu_0$ such that $\|x(\mu_n) - x(\mu_0)\| \geq \varepsilon_0$ where

$$\mathbb{E}[u(S_1 \cdot x(\mu_n))] \geq \mu_n \text{ and } \mathfrak{r}(x(\mu_n)) = \widehat{\mathfrak{r}}(\widehat{x}(\mu_n)) \leq \gamma(\mu_n). \tag{3.15}$$

By Assumption 3.15 we may assume without loss of generality that $x(\mu_n)$ converges to some portfolio $x^*$ with $\|x^* - x(\mu_0)\| \geq \varepsilon_0$. Furthermore, by Proposition 3.22 $\gamma(\mu)$ is convex and, thus, is continuous in its domain (see e.g. [18, Theorem 10.4]). Taking limits in (3.15) yields

$$\mathbb{E}[u(S_1 \cdot x^*)] \geq \mu_0 \text{ and } \widehat{\mathfrak{r}}(\widehat{x}^*) = \gamma(\mu_0). \tag{3.16}$$

But the uniqueness of the efficient portfolio (3.16) implies that $x^* = x(\mu_0)$, which is a contradiction. If $\mu_{\min}$ and/or $\mu_{\max}$ is finite and attained at some $x \in A$ then with the same arguments as above the unique continuous portfolio extends to the respective bound of $(\mu_{\min}, \mu_{\max})$.

The proof for the case when condition (c2) holds is similar. The only difference is that uniqueness of the efficient portfolio now follows from the strict concavity of the mapping $x \to \mathbb{E}[u(S_1 \cdot x)]$ (by Lemma 3.14) and the convexity of $\mathfrak{r}(x)$.

(b) We know by definition graph $\gamma(\mu)$ = graph $\nu(r)$. Moreover, $\gamma(\mu)$ is convex and, therefore, continuous on $(\mu_{\min}, \mu_{\max})$. Finally, by assumption (u1), $\gamma(\mu)$ is a strictly increasing function on $(\mu_{\min}, \mu_{\max})$. Thus $\gamma(\mu)$ is invertible on $(\mu_{\min}, \mu_{\max})$. Clearly, the inverse of $\gamma(\mu)$ is $\nu(r)$ whose corresponding domain is $(r_{\min}, r_{\max})$. The relationships $r = \gamma(\mu)$ and $\mu = \nu(r)$ characterize the pair of inverse functions $\gamma$ and $\nu$. Defining $y(r) = x(\nu(r))$ the conclusion of (b) follows.

(c) Since $A$ contains only portfolios of unit initial cost, $(1, \widehat{0})^\top \in A$ when (r1n) is satisfied. Then we can directly verify the conclusion in (c). Q.E.D.

**Remark 3.25.** *(a) When Assumption 3.15 (b) holds, then*

$$r_{\min} = \min\{\mathfrak{r}(x) : x \in A\}$$

*and*

$$\mu_{\min} = \sup\{\mathbb{E}[u(S_1 \cdot x)] : \mathfrak{r}(x) = r_{\min}, x \in A\}$$

*is also finite by (3.13). A typical efficient frontier corresponding to this case is illustrated in Figure 1.*

*(b) It is possible that $\mu_{\max}$ and/or $r_{\max}$ to be $+\infty$. Suppose $\mu_{\max}$ is finite and attained at an efficient portfolio $x(\mu_{\max})$. Under the conditions of the theorem the portfolio $\kappa := x(\mu_{\max})$ is unique and independent of the risk measure. A graphic illustration is given in Figure 3.*

*(c) Trade-off between utility and risk is thus implemented by portfolios $x(\mu)$ which trace out a curve in the leverage space of Vince [28]. Note that the curve $x(\mu)$ depends on the risk measure $\mathfrak{r}$ as well as the utility function $u$. This provides a method for systematically selecting portfolios in the leverage space to reduce risk exposure.*

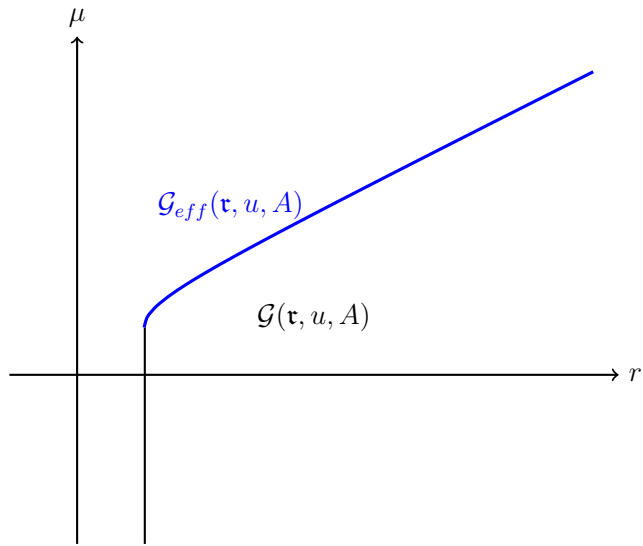

Figure 1: Efficient frontier with both $r_{\min}$ and $\mu_{\min}$ are finite and attained.

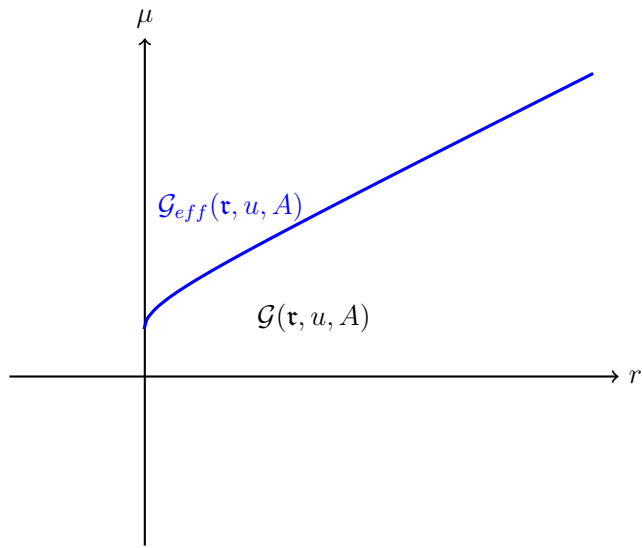

Figure 2: Efficient frontier with $(1, \widehat{0})^\top \in A$.

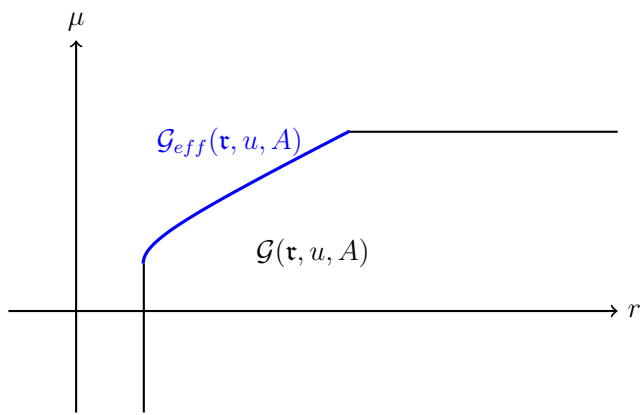

Figure 3: Efficient frontier when $r_{\min} > 0$ and $\mu_{\max}$ is finite and attained as maximum.

# 4 Markowitz Portfolio Theory and CAPM Model

Let us now turn to applications of the general theory. We show that the results in the previous section provide a general unified framework for several familiar portfolio theories. They are Markowitz portfolio theory, CAPM model, growth optimal portfolio theory and leverage space portfolio theory. Of course, when dealing with concrete risk measures and expected utilities related to these concrete theories additional helpful structure in the solutions often emerge. Although many different expositions of these theories do already exist in the literature, for convenience of readers we include brief arguments using Lagrange multiplier methods. In this entire section we will assume that the market $S_t$ from Definition 2.1 has no nontrivial riskless portfolio.

## 4.1 Markowitz Portfolio Theory

Markowitz [15] portfolio theory which considers only risky assets can be understood as a special case of the framework discussed in Section 3. The risk measure is the standard deviation $\sigma$ and the utility function is the identity function. So we face the problem

$$\begin{aligned} \min \quad & \sigma(\widehat{S}_1 \cdot \widehat{x}) \\ \text{Subject to} \quad & \mathbb{E}[\widehat{S}_1 \cdot \widehat{x}] \geq \mu, \\ & \widehat{S}_0 \cdot \widehat{x} = 1. \end{aligned} \qquad (4.1)$$

We assume $\mathbb{E}[\widehat{S}_1]$ is not proportional to $\widehat{S}_0$, that is, for any $\alpha \in \mathbb{R}$,

$$\mathbb{E}[\widehat{S}_1] \neq \alpha \widehat{S}_0. \qquad (4.2)$$

Since the variance is a monotone increasing function of the standard deviation we can minimize half of variance for convenience.

$$\min_{\widehat{x}\in\mathbb{R}^M} \quad \widehat{\mathfrak{r}}(\widehat{x}) := \frac{1}{2}\mathrm{Var}(\widehat{S}_1 \cdot \widehat{x}) = \frac{1}{2}\sigma^2(\widehat{S}_1 \cdot \widehat{x}) = \frac{1}{2}\widehat{x}^\top \Sigma \widehat{x} \qquad (4.3)$$
$$\text{Subject to} \quad \mathbb{E}[\widehat{S}_1 \cdot \widehat{x}] \geq \mu,$$
$$\widehat{S}_0 \cdot \widehat{x} = 1.$$

Optimization problem (4.3) is already in the form (3.9) with $A = \{x \in \mathbb{R}^{M+1} : S_0 \cdot x = 1, x_0 = 0\}$. We can check condition (c1) in Theorem 3.24 is satisfied. Moreover, Corollary 3.10 implies that $\Sigma$ is positive definite since $S_t$ has no nontrivial riskless portfolio. Hence, the risk function $\widehat{\mathfrak{r}}$ has compact level sets. Thus, Assumption 3.15 is satisfied and Theorem 3.24 is applicable. Let $\widehat{x}(\mu)$ be the optimal portfolio corresponding to $\mu$. Consider the Lagrangian

$$L(\widehat{x}, \lambda) := \frac{1}{2}\widehat{x}^\top \Sigma \widehat{x} + \lambda_1(\mu - \mathbb{E}[\widehat{S}_1] \cdot \widehat{x}) + \lambda_2(1 - \widehat{S}_0 \cdot \widehat{x}), \qquad (4.4)$$

where $\lambda_1 \geq 0$. Thanks for Theorem 2.9 we have

$$0 = \nabla_{\widehat{x}} L = \widehat{x}^\top(\mu)\Sigma - (\lambda_1 \mathbb{E}[\widehat{S}_1] + \lambda_2 \widehat{S}_0). \qquad (4.5)$$

In other words

$$\widehat{x}^\top(\mu) = (\lambda_1 \mathbb{E}[\widehat{S}_1] + \lambda_2 \widehat{S}_0)\Sigma^{-1}. \qquad (4.6)$$

We must have $\lambda_1 > 0$ because otherwise $\widehat{x}^\top(\mu)$ would be unrelated to the payoff $\widehat{S}_1$. The complementary slackness condition implies that $\mathbb{E}[\widehat{S}_1 \cdot \widehat{x}(\mu)] = \mu$. Right multiplying (4.5) by $\widehat{x}(\mu)$ we have

$$\sigma^2(\mu) = \lambda_1 \mu + \lambda_2. \qquad (4.7)$$

To determine the Lagrange multipliers, we need the numbers $\alpha = \mathbb{E}[\widehat{S}_1]\Sigma^{-1}\mathbb{E}[\widehat{S}_1]^\top$, $\beta = \mathbb{E}[\widehat{S}_1]\Sigma^{-1}\widehat{S}_0^\top$ and $\gamma = \widehat{S}_0 \Sigma^{-1} \widehat{S}_0^\top$. Right multiplying (4.6) by $\mathbb{E}[\widehat{S}_1]^\top$ and $\widehat{S}_0^\top$ we have

$$\mu = \lambda_1 \alpha + \lambda_2 \beta \qquad (4.8)$$

and

$$1 = \lambda_1 \beta + \lambda_2 \gamma. \qquad (4.9)$$

Solving (4.8) and (4.9) we derive

$$\lambda_1 = \frac{\gamma\mu - \beta}{\alpha\gamma - \beta^2} \quad \text{and} \quad \lambda_2 = \frac{\alpha - \beta\mu}{\alpha\gamma - \beta^2}, \qquad (4.10)$$

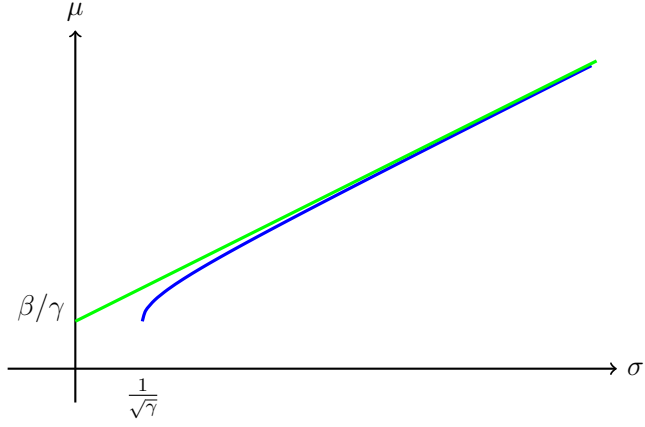

Figure 4: Markowitz Bullet

where

$$\alpha\gamma - \beta^2 = \det\left(\begin{bmatrix} \mathbb{E}[\widehat{S}_1] \\ \widehat{S}_0 \end{bmatrix} \Sigma^{-1}[\mathbb{E}[\widehat{S}_1^\top], \widehat{S}_0^\top]\right) > 0 \qquad (4.11)$$

since $\Sigma^{-1}$ is positive definite and condition (4.2) holds. Substituting (4.10) into (4.7) we see that the efficient frontier is determined by the curve

$$\sigma(\mu) = \sqrt{\frac{\gamma\mu^2 - 2\beta\mu + \alpha}{\alpha\gamma - \beta^2}} = \sqrt{\frac{\gamma}{\alpha\gamma - \beta^2}\left(\mu - \frac{\beta}{\gamma}\right)^2 + \frac{1}{\gamma}} \geq \frac{1}{\sqrt{\gamma}} \qquad (4.12)$$

usually referred to as the Markowitz bullet due to its shape. A typical Markowitz bullet is shown in Figure 4 with an asymptote

$$\mu = \frac{\beta}{\gamma} + \sigma(\mu)\sqrt{\frac{\alpha\gamma - \beta^2}{\gamma}}. \qquad (4.13)$$

Note that $\mathcal{G}(\frac{1}{2}\mathrm{Var}, id, \{S_0 \cdot x = 1, x_0 = 0\}) = \mathcal{G}(\sigma, id, \{S_0 \cdot x = 1, x_0 = 0\})$. Thus, relationships (4.12) and (4.13) describe the efficient frontier $\mathcal{G}_{eff}(\sigma, id, \{S_0 \cdot x = 1, x_0 = 0\})$ as in Definition 3.18. Also note that (4.12) implies that $\mu_{\min} = \beta/\gamma$ and $r_{\min} = 1/\sqrt{\gamma}$. Thus, as a corollary of Theorem 3.24, we have

**Theorem 4.1.** (Markowitz Portfolio Theorem) *Assume that the financial market $S_t$ has no nontrivial riskless portfolio and $\mathbb{E}[\widehat{S}_1]$ is not proportional to $\widehat{S}_0$ (see (4.2)). The Markowitz efficient portfolios of (4.1) represented in the $(\sigma, \mu)-$plane are given by*

$$\mathcal{G}_{eff}(\sigma, id; \{S_0 \cdot x = 1, x_0 = 0\}).$$

*They correspond to the upper boundary of the Markowitz bullet given by*

$$\sigma(\mu) = \sqrt{\frac{\gamma\mu^2 - 2\beta\mu + \alpha}{\alpha\gamma - \beta^2}}, \ \mu \in \left[\frac{\beta}{\gamma}, +\infty\right).$$

*The optimal portfolio $\widehat{x}(\mu)$ can be determined by (4.6) and (4.10) as*

$$\widehat{x}(\mu) = \mu\frac{\Sigma^{-1}(\gamma\mathbb{E}[\widehat{S}_1^\top] - \beta\widehat{S}_0^\top)}{\alpha\gamma - \beta^2} + \frac{\Sigma^{-1}(\alpha\widehat{S}_0^\top - \beta\mathbb{E}[\widehat{S}_1^\top])}{\alpha\gamma - \beta^2}, \tag{4.14}$$

*which is affine in $\mu$.*

The structure of the optimal portfolio in (4.14) implies the well known two fund theorem derived by Tobin in [26].

**Theorem 4.2.** (Two Fund Theorem) *Select two distinct portfolios on the Markowitz efficient frontier. Then any portfolio on the Markowitz efficient frontier can be represented as the linear combination of these two portfolios.*

**Remark 4.3.** The two fund theorem can be viewed as the theoretical foundation for the passive investment strategy of buy and hold broad based indices. Since most mutual funds and hedge funds underperform the broad based indices, empirically we can regard broad based indices such as SP500 and NASDAQ as Markowitz efficient portfolios. By the two fund theorem holding two such broad based indices passively we can produce any efficient portfolio on the Markowitz bullet.

## 4.2 Capital Asset Pricing Model

The capital asset pricing model (CAPM) is a theoretical model independently proposed by Lintner [9], Mossin [17], Sharpe [22] and Treynor [25] for pricing a risky asset according to its expected payoff and market risk, often referred to as the beta. The core of the capital asset pricing model is an extension of the Markowitz portfolio theory to include a riskless bond. Thus we can apply the general framework in Section 3 with the same setting as in Section 4.1. Similar to the previous section we can consider the equivalent problem of

$$\begin{aligned}
\min_{x \in \mathbb{R}^{M+1}} \quad & \frac{1}{2}\sigma^2(S_1 \cdot x) = \frac{1}{2}\widehat{x}^\top \Sigma \widehat{x} =: \widehat{\mathfrak{r}}(\widehat{x}) \\
\text{Subject to} \quad & \mathbb{E}[S_1 \cdot x] \geq \mu, \\
& S_0 \cdot x = 1.
\end{aligned} \tag{4.15}$$

Similar to the last section problem (4.15) is in the form (3.9) with $A = \{x \in \mathbb{R}^{M+1} : S_0 \cdot x = 1\}$. We can check condition (c1) in Theorem 3.24 is satisfied. Again the risk function $\widehat{\mathfrak{r}}$ has compact level sets since $\Sigma$ is positive definite. Thus, Assumption 3.15 is satisfied and Theorem 3.24 is applicable. The Lagrangian of this convex programming problem is

$$L(x,\lambda) := \frac{1}{2}\widehat{x}^\top \Sigma \widehat{x} + \lambda_1(\mu - \mathbb{E}[S_1] \cdot x) + \lambda_2(1 - S_0 \cdot x), \tag{4.16}$$

where $\lambda_1 \geq 0$. Again we have

$$0 = \nabla_x L = (0, \widehat{x}^\top(\mu)\Sigma) - (\lambda_1 \mathbb{E}[S_1] + \lambda_2 S_0). \tag{4.17}$$

Using $S_1^0 = R$ and $S_0^0 = 1$, the first component of (4.17) implies

$$\lambda_2 = -\lambda_1 R. \tag{4.18}$$

So that (4.17) becomes

$$0 = \nabla_x L = (0, \widehat{x}^\top(\mu)\Sigma) - \lambda_1(\mathbb{E}[S_1] - R S_0). \tag{4.19}$$

Clearly $\lambda_1 > 0$ for $\widehat{x}(\mu) \neq 0$. Using the complementary slackness condition $\mathbb{E}[S_1 \cdot x(\mu)] = \mu$ we derive

$$\sigma^2(\mu) = \widehat{x}^\top(\mu) \Sigma \widehat{x}(\mu) = \lambda_1(\mu - R), \tag{4.20}$$

by right multiplying $x(\mu)$ in (4.19). Solving $\widehat{x}^\top(\mu)$ from (4.19) we have

$$\widehat{x}^\top(\mu) = \lambda_1(\mathbb{E}[\widehat{S}_1] - R\widehat{S}_0)\Sigma^{-1}. \tag{4.21}$$

Right multiplying with $\mathbb{E}[\widehat{S}_1^\top]$ and $\widehat{S}_0^\top$ and using the $\alpha, \beta$ and $\gamma$ introduced in the previous section we derive

$$\mu - x_0(\mu)R = \lambda_1(\alpha - R\beta) \tag{4.22}$$

and

$$1 - x_0(\mu) = \lambda_1(\beta - R\gamma), \tag{4.23}$$

respectively. Multiplying (4.23) by $R$ and subtract it from (4.22) we get

$$\mu - R = \lambda_1(\alpha - 2\beta R + \gamma R^2). \tag{4.24}$$

Combining (4.20) and (4.24) we arrive at

$$\sigma^2(\mu) = \frac{(\mu - R)^2}{\alpha - 2\beta R + \gamma R^2}. \tag{4.25}$$

It only makes sense to involve risky assets when we can expect an excess return. Thus, $\mu \geq R$. Relation (4.25) defines a straight line on the $(\sigma, \mu)$-plane

$$\sigma(\mu) = \frac{\mu - R}{\sqrt{\Delta}} \quad \text{or} \quad \mu = R + \sigma(\mu)\sqrt{\Delta}, \tag{4.26}$$

where $\Delta := \alpha - 2\beta R + \gamma R^2 > 0$ if

$$\mathbb{E}[\widehat{S}_1] - R\widehat{S}_0 \neq 0 \tag{4.27}$$

since $\Sigma$ is positive definite. The line given in (4.26) is called the *capital market line*.

Also combining (4.21), (4.23) and (4.24) we have

$$x^\top(\mu) = \Delta^{-1}[\alpha - \beta R - \mu(\beta - \gamma R), (\mu - R)(\mathbb{E}[\widehat{S}_1] - R\widehat{S}_0)\Sigma^{-1}]. \tag{4.28}$$

Again we see the affine structure of the solution. In particular, when $\mu = R$ and $\mu = (\alpha - \beta R)/(\beta - \gamma R)$ we derive, respectively, the portfolio $(1, \widehat{0})^\top$ that contains only the riskless bond and the portfolio $(0, (\mathbb{E}[\widehat{S}_1] - R\widehat{S}_0)\Sigma^{-1}/(\beta - \gamma R))^\top$ that contains only risky assets. We call this portfolio the *market portfolio* and denote it $x_M$. The market portfolio corresponds to the coordinates

$$(\sigma_M, \mu_M) = \left(\frac{\sqrt{\Delta}}{\beta - \gamma R}, R + \frac{\Delta}{\beta - \gamma R}\right). \tag{4.29}$$

Since the risk $\sigma$ is non negative we see that the market portfolio exists only when

$$\beta - \gamma R > 0.$$

This condition is

$$(\mathbb{E}[\widehat{S}_1] - R\widehat{S}_0) \cdot \Sigma^{-1}\widehat{S}_0^\top > 0. \tag{4.30}$$

Note that (4.30) also implies (4.27).

Again note that although the computation is done in terms of the risk function $\widehat{\mathfrak{r}}(\widehat{x}) = \frac{1}{2}\widehat{x}^\top \Sigma \widehat{x}$, relationships in (4.26) are in terms the risk function $\sigma(S_1 \cdot x)$. Thus, they describe the efficient frontier $\mathcal{G}_{eff}(\sigma, id, S_0 \cdot x = 1)$ as in Definition 3.18. In summary, we have

**Theorem 4.4.** (CAPM) *Assume that the financial market $S_t$ of Definition 2.1 has no nontrivial riskless portfolio. Moreover assume that condition (4.30) holds. The efficient portfolios for the CAPM model $\mathcal{G}_{eff}(\sigma, id; \{S_0 \cdot x = 1\})$ represented in the $(\sigma, \mu)$-plane are a straight line passing through $(0, R)$ corresponding to the portfolio of pure risk free bond and $(\sigma_M, \mu_M)$ corresponding to the market portfolio of purely risky assets. The optimal portfolio $x(\mu)$ can be determined by (4.28) which is affine in $\mu$.*

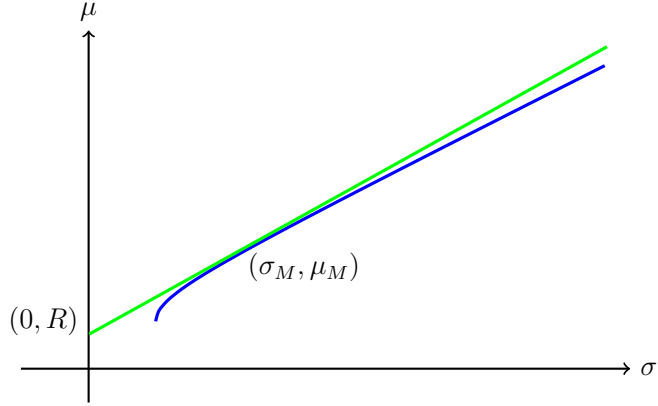

Figure 5: Capital Market Line and Markowitz Bullet

By Theorem 3.20

$$(\sigma_M, \mu_M) \in \mathcal{G}_{eff}(\sigma, id; \{S_0 \cdot x = 1\}) \cap \mathcal{G}(\sigma, id; \{S_0 \cdot x = 1, x_0 = 0\}) \quad (4.31)$$
$$\subset \mathcal{G}_{eff}(\sigma, id; \{S_0 \cdot x = 1, x_0 = 0\}).$$

Thus, the market portfolio has to reside on the Markowitz efficient frontier. Moreover, by (4.28) we can see that the market portfolio $x_M$ is the only portfolio on the CAPM efficient frontier that consists of purely risky assets. Thus,

$$\mathcal{G}_{eff}(\sigma, id; \{S_0 \cdot x = 1\}) \cap \mathcal{G}(\sigma, id; \{S_0 \cdot x = 1, x_0 = 0\}) = \{(\sigma_M, \mu_M)\}, \quad (4.32)$$

so that the capital market line is tangent to the Markowitz bullet at $(\sigma_M, \mu_M)$ as illustrated in Figure 5. The affine structure of the solutions is summarized in the following one fund theorem [22, 26].

**Theorem 4.5.** (One Fund Theorem) *Assume that the financial market $S_t$ has no non-trivial riskless portfolio. Moreover assume that condition (4.30) holds. All the optimal portfolios in the CAPM model (4.15) are generalized convex combinations of the riskless bond and the market portfolio $x_M = (0, (\mathbb{E}[\widehat{S}_1] - R\widehat{S}_0)\Sigma^{-1}/(\beta - \gamma R))^\top$. Optimal portfolios $x(\mu)$ are affine in $\mu$ (see (4.28)) and can be represented as points in the $(\sigma, \mu)$-plane as located on the* capital market line

$$\mu = R + \sigma\sqrt{\Delta}, \ \sigma \geq 0.$$

*The capital market line is tangent to the boundary of the Markowitz bullet at the coordinates of the market portfolio $(\sigma_M, \mu_M)$ and intercepts the $\mu$ axis at $(0, R)$ (see Fig. 5).*

**Remark 4.6.** *The one fund theorem combined with the two fund theorem provides a theoretical foundation for the passive investment strategy. The two fund theorem implies* *that if two broad based indices are approximately on the Markowitz frontier then we can use a linear combination of these two indices to derive the market portfolio. Thus, by the one fund theorem in order to construct an efficient portfolio in the sense of the CAPM model we only need to consider a mix of the bond and the two indices.*

Alternatively we can write the slope of the capital market line as

$$\sqrt{\Delta} = \frac{\mu_M - R}{\sigma_M}. \qquad (4.33)$$

This quantity is called the *price of risk* and we can rewrite the equation for the capital market line (4.26) as

$$\mu = R + \frac{\mu_M - R}{\sigma_M}\sigma. \qquad (4.34)$$

**Remark 4.7.** (Sharpe Ratio) *We note that for any given portfolio $x$ its corresponding pair of coordinates $(\sigma, \mu)$ in the risk-return space also produces a ratio*

$$\frac{\mu - R}{\sigma}. \qquad (4.35)$$

*In the risk-return space this is the slope of the line representing portfolios mixing $x$ with a riskless bond. Clearly the larger this ratio the better the portfolio serves this purpose. Sharpe [23] proposed to use this ratio, later called Sharpe ratio, to measure the performance of mutual funds.*

We can also use the capital market line to price a risky asset as we initially set out to do. The pricing principle in the capital asset pricing model is that adding a fair priced risky asset to the market should not change the capital market line. For convenience we assume that the price is implied by the expected return of the asset. Thus, given a risky asset $a^i$, we try to determine its expected return $\mu_i$.

**Theorem 4.8.** (Capital Asset Pricing Model: the beta) *Assume that the financial market $S_t$ of Definition 2.1 has no nontrivial riskless portfolio. Moreover assume that condition (4.30) holds. Let $a_0^i$ be the fair price of a risky asset $a^i$ with a payoff $a_1^i$ at $t = 1$. Denote the expected percentage return of $a^i$ by $\mu_i = \mathbb{E}[a_1^i]/a_0^i$. Then*

$$\mu_i = R + \beta_i(\mu_M - R). \qquad (4.36)$$

*Here $\beta_i := \sigma_{iM}/\sigma_M^2$ is called the beta of $a^i$, where $\sigma_{iM} := \text{Cov}(a_1^i/a_0^i, S_1 \cdot x_M)$ is the covariance of $a_1^i/a_0^i$ and the payoff of the market portfolio.*

**Proof.** Consider a portfolio relies on the parameter $\alpha$ that mixes the risky asset $a^i$ and the market portfolio:

$$p(\alpha) = \alpha a_1^i/a_0^i + (1 - \alpha)S_1 \cdot x_M. \qquad (4.37)$$

Denote the expected return and the standard deviation of $p(\alpha)$ by $\mu_\alpha$ and $\sigma_\alpha$, respectively. Hence we have

$$\mu_\alpha = \alpha \mu_i + (1-\alpha)\mu_M, \tag{4.38}$$

and

$$\sigma_\alpha^2 = \alpha^2 \sigma_i^2 + 2\alpha(1-\alpha)\sigma_{iM} + (1-\alpha)^2 \sigma_M^2, \tag{4.39}$$

where $\sigma_i^2$ is the variance of $a_1^i/a_0^i$. The parametric curve $(\sigma_\alpha, \mu_\alpha)$ must lie below the capital market line because the latter consists of optimal portfolios. On the other hand it is clear that when $\alpha = 0$ this curve coincides with the capital market line. Thus, the capital market line is tangent to the line of the parametric curve $(\sigma_\alpha, \mu_\alpha)$ at $\alpha = 0$. Since the slope of the capital market line is $(\mu_M - R)/\sigma_M$, it follows that

$$\frac{\mu_M - R}{\sigma_M} = \left[\frac{d\mu_\alpha}{d\sigma_\alpha}\right]_{\alpha=0} = \frac{\sigma_M(\mu_i - \mu_M)}{\sigma_{iM} - \sigma_M^2}. \tag{4.40}$$

Solving for $\mu_i$ we derive

$$\mu_i = R + \beta_i(\mu_M - R). \tag{4.41}$$

Q.E.D.

## 5 Affine Structure of the Efficient Portfolios

The affine dependence of the efficient portfolio on the return $\mu$ observed in the CAPM still holds when the standard deviation is replaced by the more general deviation measure (see [20]. In this section we derive this affine structure using the general framework discussed in Section 3 and provide a proof different from that of [20]. We also construct a counterexample showing that the two fund theorem (Theorem 4.2) fails in this setting. Let's consider a risk measure $\mathfrak{r}$ that satisfies (r1), (r1n), (r2) and (r3) in Assumption 3.1 and the related problem of finding efficient portfolios becomes

$$\begin{aligned}
\min_{x \in \mathbb{R}^{M+1}} \quad & \mathfrak{r}(x) = \widehat{\mathfrak{r}}(\widehat{x}) \\
\text{Subject to} \quad & \mathbb{E}[S_1 \cdot x] \geq \mu, \\
& S_0 \cdot x = 1.
\end{aligned} \tag{5.1}$$

Since for $\mu = R$ there is an obvious solution $x(R) = (1, \widehat{0})$ corresponding to $\mathfrak{r}(x(R)) = \widehat{\mathfrak{r}}(\widehat{0}) = 0$, we have $r_{\min} = 0$ and $\mu_{\min} = R$. In what follows we will only consider $\mu > R$. Moreover, we note that for $\widehat{\mathfrak{r}}$ satisfying the positive homogeneous property (r3) in Assumption 3.1, $\widehat{y} \in \partial \widehat{\mathfrak{r}}(\widehat{x})$ implies that

$$\widehat{\mathfrak{r}}(\widehat{x}) = \langle \widehat{y}, \widehat{x} \rangle. \tag{5.2}$$

In fact, for any $t \in (-1, 1)$,

$$t\widehat{\mathfrak{r}}(\widehat{x}) = \widehat{\mathfrak{r}}((1+t)\widehat{x}) - \widehat{\mathfrak{r}}(\widehat{x}) \geq t\langle \widehat{y}, \widehat{x}\rangle, \tag{5.3}$$

and (5.2) follows. Now we can state and prove the theorem on affine dependence of the efficient portfolio on the return $\mu$.

**Theorem 5.1.** (Affine Efficient Frontier for Positive Homogeneous Risk Measures) *Assume that the financial market $S_t$ of Definition 2.1 has no nontrivial riskless portfolio. Assume that the risk measure $\mathfrak{r}$ satisfies assumptions (r1), (r1n), (r2) and (r3) in Assumption 3.1 with $A = \{x \in \mathbb{R}^{M+1} : S_0 \cdot x = 1\}$ and Assumption 3.15 (b) holds. Furthermore, assume that there exists some $\bar{m} \in \{1, 2, \ldots, M\}$ with*

$$\mathbb{E}[S_1^{\bar{m}}] \neq R S_0^{\bar{m}}. \tag{5.4}$$

*Then there exists an efficient portfolio $x^1$ corresponding to $(r_1, \mu_1) = (\mathfrak{r}(x^1), R + 1)$ on the efficient frontier for problem (5.1) such that the efficient frontier for problem (5.1) in the risk-expected return space is a straight line that passes through the points (0,R) corresponding to a portfolio of pure bond $(1, \widehat{0})^\top$ and $(r_1, \mu_1)$ corresponding to the portfolio $x_1$, respectively. Moreover, the straight line connecting $(1, \widehat{0})^\top$ and $x_1$ in the portfolio space, namely for $\mu \geq R$,*

$$(\mu_1 - \mu)(1, \widehat{0})^\top + (\mu - R)x^1 \tag{5.5}$$

*represents a set of efficient portfolios that corresponds to*

$$(\gamma(\mu), \mu) = ((\mu - R)r_1, \mu) \tag{5.6}$$

*in the risk-expected return space (see Definition 3.18 and (3.9)).*

**Proof.** The Lagrangian of this convex programming problem (5.1) is

$$L(x, \lambda) := \mathfrak{r}(x) + \lambda_1(\mu - \mathbb{E}[S_1] \cdot x) + \lambda_2(1 - S_0 \cdot x), \tag{5.7}$$

where $\lambda_1 \geq 0$ and $\lambda_2 \in \mathbb{R}$.

Condition (5.4) implies that, for any $\mu$ there exists a portfolio of the form $y = (y_0, 0, \ldots, 0, y_{\bar{m}}, 0, \ldots, 0)^\top$ satisfying

$$\begin{bmatrix} \mathbb{E}[S_1 \cdot y] \\ S_0 \cdot y \end{bmatrix} = \begin{bmatrix} Ry_0 + \mathbb{E}[S_1^{\bar{m}}]y_{\bar{m}} \\ y_0 + S_0^{\bar{m}} y_{\bar{m}} \end{bmatrix} = \begin{bmatrix} R & \mathbb{E}[S_1^{\bar{m}}] \\ 1 & S_0^{\bar{m}} \end{bmatrix} \begin{bmatrix} y_0 \\ y_{\bar{m}} \end{bmatrix} = \begin{bmatrix} \mu \\ 1 \end{bmatrix}, \tag{5.8}$$

because the matrix in (5.8) is invertible. Thus, for any $\mu \geq R$, Assumption 3.15 (b) with $A = \{x \in \mathbb{R}^{M+1} : S_0 \cdot x = 1\}$ and condition (5.4) ensure the existence of an optimal solution to problem (5.1).

Denoting one of those solutions by $x(\mu)$ (may not be unique) we have

$$\gamma(\mu) = \mathfrak{r}(x(\mu)) = \widehat{\mathfrak{r}}(\widehat{x}(\mu)). \tag{5.9}$$

Fixing $\mu_1 = R + 1 > R$, denote $x^1 = x(\mu_1)$. Then

$$\lambda_1 \mathbb{E}[S_1] + \lambda_2 S_0 \in \partial \mathfrak{r}(x^1). \tag{5.10}$$

Since $\mathfrak{r}$ is independent of $x_0$ we have

$$\lambda_1 \mathbb{E}[S_1^0] + \lambda_2 S_0^0 = 0 \text{ or } \lambda_2 = -\lambda_1 R. \tag{5.11}$$

Substituting (5.11) into (5.10) we have

$$\lambda_1 \mathbb{E}[\widehat{S}_1 - R\widehat{S}_0] \in \partial \widehat{r}(\widehat{x}^1) \tag{5.12}$$

so that, for all $\widehat{x} \in \mathbb{R}^M$,

$$\widehat{\mathfrak{r}}(\widehat{x}) - \widehat{\mathfrak{r}}(\widehat{x}^1) \geq \lambda_1 \mathbb{E}[(\widehat{S}_1 - R\widehat{S}_0) \cdot (\widehat{x} - \widehat{x}^1)] = \lambda_1 (\mathbb{E}[(\widehat{S}_1 - R\widehat{S}_0) \cdot \widehat{x}] - (\mu_1 - R)) \tag{5.13}$$

because at the optimal solution $\widehat{x}^1$ the constraint is binding. Using (r3) it follows from (5.2) and (5.12) that

$$\widehat{\mathfrak{r}}(\widehat{x}^1) = \lambda_1 \mathbb{E}[(\widehat{S}_1 - R\widehat{S}_0) \cdot \widehat{x}^1] = \lambda_1(\mu_1 - R) = \lambda_1. \tag{5.14}$$

Thus, we can write (5.13) as

$$\widehat{\mathfrak{r}}(\widehat{x}) \geq \widehat{\mathfrak{r}}(\widehat{x}^1) \mathbb{E}[(\widehat{S}_1 - R\widehat{S}_0) \cdot \widehat{x}]. \tag{5.15}$$

For $t \geq 0$ define the homotopy between $x^0 := (1, \widehat{0})^\top$ and $x^1$

$$x^t := (tx_0^1 + (1-t), t\widehat{x}^1). \tag{5.16}$$

We can verify that $S_0 \cdot x^t = 1$ and

$$\mathbb{E}[S_1 \cdot x^t] = R + t$$

so that

$$\mathbb{E}[(S_1 - RS_0) \cdot x^t] = t. \tag{5.17}$$

On the other hand it follows from assumptions (r1) and (r3) that

$$\mathfrak{r}(x^t) = \widehat{\mathfrak{r}}(t\widehat{x}^1) = t\,\widehat{\mathfrak{r}}(\widehat{x}^1). \tag{5.18}$$

Thus, for any $x$ satisfying $S_0 \cdot x = 1$ and

$$\mathbb{E}[S_1 \cdot x] \geq R + t$$

it follows from (5.15) that

$$\widehat{\mathfrak{r}}(\widehat{x}) \geq \widehat{\mathfrak{r}}(\widehat{x}^1)t. \tag{5.19}$$

For any $\mu > R$, letting $t_\mu := \mu - R$, we have $\mu = R + t_\mu$. Thus, by inequality (5.19) we have $\widehat{\mathfrak{r}}(\widehat{x}(\mu)) \geq t_\mu \widehat{\mathfrak{r}}(\widehat{x}^1)$. On the other hand $x(\mu)$ is an efficient portfolio implies that $\widehat{\mathfrak{r}}(\widehat{x}(\mu)) \leq \widehat{\mathfrak{r}}(\widehat{x}^{t_\mu}) = t_\mu \widehat{\mathfrak{r}}(\widehat{x}^1)$ yielding equality

$$\gamma(\mu) = \widehat{\mathfrak{r}}(\widehat{x}(\mu)) = \widehat{\mathfrak{r}}(\widehat{x}^{t_\mu}) = t_\mu \widehat{\mathfrak{r}}(\widehat{x}^1) = (\mu - R)\widehat{\mathfrak{r}}(\widehat{x}^1). \tag{5.20}$$

In other words $\gamma(\mu)$ is an affine function in $\mu$. Also, we conclude that points $(\gamma(\mu), \mu)$ on this efficient frontier correspond to efficient portfolios

$$x^{t_\mu} = ((\mu - R)x_0^1 + \mu_1 - \mu, (\mu - R)\widehat{x}^1) = (\mu_1 - \mu)(1, \widehat{0})^\top + (\mu - R)x_1 \tag{5.21}$$

as an affine mapping of the parameter $\mu$ into the portfolio space.

Also using $r_1$ we can write (5.20) as

$$\gamma(\mu) = r_1(\mu - R). \tag{5.22}$$

That is to say the efficient frontier of (5.1) in the risk-expected return space is given by the parameterized straight line (5.6). Q.E.D.

**Remark 5.2.** (a) Clearly, $x^{t_R}$ corresponds to the portfolio $(1, \widehat{0})^\top$ with $\gamma(R) = \widehat{\mathfrak{r}}(\widehat{0}) = 0$. If $x_0^1 \neq 1$. Setting $\mu_M := \frac{\mu_1 - Rx_0^1}{1 - x_0^1}$ and $r_M := \gamma(\mu_M) = \widehat{\mathfrak{r}}(\widehat{x}^1)/(1 - x_0^1)$ we see that $(r_M, \mu_M)$ on the efficient frontier corresponds to a purely risky efficient portfolio of (5.1)

$$x_M := x^{t_{\mu_M}} = \left(0, \frac{1}{1 - x_0^1} \widehat{x}^1\right)^\top. \tag{5.23}$$

Since $x_M$ belongs to the image of the affine mapping in (5.21), the family of efficient portfolios as described by the affine mapping in (5.21) contains both the pure bond $(1, \widehat{0})^\top$ and the portfolio $x_M$ that consists only of purely risky assets. In fact, we can represent the affine mapping in (5.21) as a parametrized line passing through $(1, \widehat{0})^\top$ and $x_M$ as

$$x^{t_\mu} = \left(1 - \frac{\mu - R}{\mu_M - R}\right)(1, \widehat{0})^\top + \frac{\mu - R}{\mu_M - R} x_M, \tag{5.24}$$

which is a similar representation of the efficient portfolios as (5.5). The portfolio $x_M$ is called a *master fund* in [20]. When $\mathfrak{r} = \sigma$ it is the *market portfolio* in the CAPM. For a general risk measure $\mathfrak{r}$ satisfying conditions (r1), (r1n), (r2) and (r3) in Assumption 3.1 the master funds $x_M$ are not necessarily unique. However, all master funds correspond to the same point $(r_M, \mu_M)$ in the risk-expected return space.

(b) We can also consider problem (5.1) on the set of admissible portfolios of purely risky assets, namely $\mathcal{G}_{eff}(\mathfrak{r}, id, \{S_0 \cdot x = 1, x_0 = 0\})$. Then similar to the relationship between the Markowitz efficient frontier and the capital market line, it follows from Theorem 5.1 that

$$\mathcal{G}_{eff}(\mathfrak{r}, id, \{S_0 \cdot x = 1, x_0 = 0\}) \cap \mathcal{G}_{eff}(\mathfrak{r}, id, \{S_0 \cdot x = 1\}) = \{(r_M, \mu_M)\}, \tag{5.25}$$

*as illustrated in Figure 6.*

*(c) If $x_0^1 = 1$ then the efficient portfolios in (5.5) are related to $\mu$ in a much simpler fashion*

$$(1, \widehat{0})^\top + (\mu - R)\widehat{x}^1. \tag{5.26}$$

*In this case there is no master fund as observed in [20]. In the language of [20], portfolio $x^1$ is called a* basic fund. *Thus, Theorem 5.1 recovers the results in Theorem 2 and Theorem 3 in [20] with a different proof and a weaker condition (condition (5.4) is weaker than (A2) on page 752 of Rockafellar et al [20]).*

Since the standard deviation satisfies Assumptions (r1), (r1n), (r2) and (r3), the result above is a generalization of the relationship between the CAPM model and the Markowitz portfolio theory. We note that the standard deviation is not the only risk measure that satisfies these assumptions. For example, some forms of approximation to the expected drawdowns also satisfy these assumptions (cf. [14]).

Theorem 5.1 is a full generalization of the one fund theorem (Theorem 4.5) in the previous section. On the other hand it has been noted in footnote 10 in [20] that a similar generalization of the two fund theorem (Theorem 4.2) is not to be expected. We construct a concrete counter-example below.

**Example 5.3.** (Counter-example to a Generalized Two Fund Theorem) *Let's consider for example*

$$\min_{\widehat{x} \in \mathbb{R}^3} \quad \widehat{\mathfrak{r}}(\widehat{x}) \tag{5.27}$$
$$\text{Subject to} \quad \mathbb{E}[\widehat{S}_1 \cdot \widehat{x}] \geq \mu,$$
$$\widehat{S}_0 \cdot \widehat{x} = 1,$$

*with $M = 3$.*

*Choose all $S_0^m = 1$, so that $\widehat{S}_0 \cdot \widehat{x} = 1$ is $x_1 + x_2 + x_3 = 1$. Choose the payoff $S_1$ such that $\mathbb{E}[\widehat{S}_1 \cdot \widehat{x}] = x_1$ so that $x_1 = \mu$ at the optimal solution. Finally, let's construct $\widehat{\mathfrak{r}}(\widehat{x})$ so that the optimal solution $\widehat{x}(\mu)$ is not affine in $\mu$.*

*We do so by constructing a convex set $G$ with $0 \in \text{int}G$ (interior of $G$) and then set $\widehat{\mathfrak{r}}(\widehat{x}) = 1$ for $\widehat{x} \in \partial G$ (boundary of $G$) and extend $\widehat{\mathfrak{r}}$ to be positive homogeneous. Then (r1), (r1n), (r2) and (r3) are satisfied.*

*Now let's specify $G$. Take the convex hull of the set $[-5, 5] \times [-1, 1] \times [-1, 1]$ and five other points. One point is $E = (10, 0, 0)^\top$ and the other four points $A, B, C$ and $D$, are the corner points of a square that lies in the plane $x_1 = 9$ and has unit side length. To obtain that square take the standard square with unit side length in $x_1 = 9$, i.e. the square with corner points $(9, \pm 1/2, \pm 1/2)^\top$ and rotate this square by 30 degrees counter*

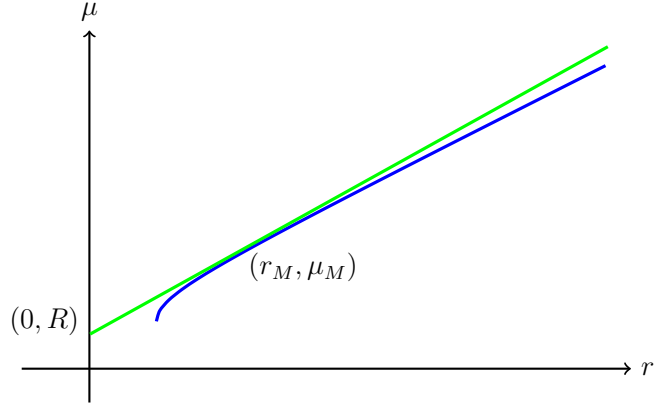

Figure 6: Capital Market Line for (5.1) when $x_0^1 \neq 1$

*clockwise in the $x_2 x_3$-plane. Doing some calculation one gets:*

$$
\begin{aligned}
A &= (9, (-1+\sqrt{3})/4, (1+\sqrt{3})/4)^\top \\
B &= (9, (-1-\sqrt{3})/4, (-1+\sqrt{3})/4)^\top \\
C &= (9, (1-\sqrt{3})/4, -(1+\sqrt{3})/4))^\top \\
D &= (9, (1+\sqrt{3})/4), (1-\sqrt{3})/4))^\top.
\end{aligned}
$$

*Obviously for $\mu = 1$ the optimal solution is $\widehat{x}(1) = (1,0,0)^\top$ with $\widehat{\mathfrak{r}}(\widehat{x}(1)) = 1/10$ For $\mu = 1+\epsilon$ with $\epsilon > 0$ small we have $\widehat{x}(1+\epsilon) = (1+\epsilon, \epsilon\sqrt{3}(+1-\sqrt{3})/6, \epsilon\sqrt{3}(-1-\sqrt{3})/6))^\top$ (they lie on the ray through a point on the convex combination of $C$ and $(10,0,0)^\top$) and for $\mu = 1+d$ with $d > 0$ large we have $\widehat{x}(1+d) = (1+d, -d/2, -d/2)^\top$ (they lie on the ray through a point on the set $\{(x_1, -1, -1)^\top : x_1 \in (2,5)\}$. Therefore, $\widehat{x}(\mu)$ cannot be affine in $\mu$.*

# 6 Growth Optimal and Leverage Space Portfolio

Growth portfolio theory is proposed by Lintner [9] and is also related to the work of Kelly [8]. It is equivalent to maximizing the expected log utility:

$$
\begin{aligned}
\max_{x \in \mathbb{R}^{M+1}} \quad & \mathbb{E}[\ln(S_1 \cdot x)] \\
\text{Subject to} \quad & S_0 \cdot x = 1.
\end{aligned}
\tag{6.1}
$$

**Remark 6.1.** *Problem (6.1) is equivalent to*

$$
\max_{\widehat{x} \in \mathbb{R}^M} \mathbb{E}[\ln(R + (\widehat{S}_1 - R\widehat{S}_0) \cdot \widehat{x})]
\tag{6.2}
$$

**Theorem 6.2.** (Growth Optimal Portfolio) *Assume that the financial market $S_t$ of Definition 2.1 has no nontrivial riskless portfolio. Then problem (6.1) has a unique optimal portfolio, which is often referred to as the* growth optimal portfolio *and is denoted $\kappa \in \mathbb{R}^{M+1}$.*

To prove Theorem 6.2 we need the following lemma.

**Lemma 6.3.** *Assume that the financial market $S_t$ of Definition 2.1 has no nontrivial riskless portfolio. Let $u$ be a continuous utility function satisfying (u3) in Assumption 3.3. Then for any $\mu \in \mathbb{R}$,*

$$\{x \in \mathbb{R}^{M+1} : \mathbb{E}[u(S_1 \cdot x)] \geq \mu, S_0 \cdot x = 1\} \tag{6.3}$$

*is compact (and possibly empty in some cases).*

**Proof.** Since $u$ is continuous, the set in (6.3) is closed. Thus, we need only to show it is also bounded. Assume the contrary that there exists a sequence of portfolios $x^n$ with

$$S_0 \cdot x^n = 1 \tag{6.4}$$

and $\|x^n\| \to \infty$ satisfying

$$\mathbb{E}[u(S_1 \cdot x^n)] \geq \mu. \tag{6.5}$$

Equation (6.4) implies that $\|\widehat{x}^n\| \to \infty$. Then without loss of generality we may assume $x^n/\|\widehat{x}^n\|$ converges to $x^* = (x_0^*, \widehat{x}^*)^\top$ where $\|\widehat{x}^*\| = 1$. Condition (u3) and (6.5) for arbitrary $\mu \in \mathbb{R}$ imply that, for each natural number $n$,

$$S_1 \cdot x^n \geq 0. \tag{6.6}$$

Dividing (6.4) and (6.6) by $\|\widehat{x}^n\|$ and taking limits as $n \to \infty$ we derive

$$S_0 \cdot x^* = 0 \tag{6.7}$$

and

$$S_1 \cdot x^* \geq 0. \tag{6.8}$$

Combining (6.7) and (6.8) we have

$$(\widehat{S}_1 - R\widehat{S}_0) \cdot \widehat{x}^* \geq 0, \tag{6.9}$$

and thus $x^*$ is a nontrivial riskless portfolio, which is a contradiction. Q.E.D.

**Proof. of Theorem 6.2** We can verify that the utility function $u = \ln$ satisfies conditions (u1), (u2s), (u3) and (u4). Also $\{x : \mathbb{E}[\ln(S_1 \cdot x)] \geq \ln(R), S_0 \cdot x = 1\} \neq \emptyset$ because it contains $(1, \widehat{0})^\top$. Thus, Lemma 6.3 implies that problem (6.1) has at least one solution and
$$\mu_{\max} = \max_{x \in \mathbb{R}^{M+1}} \{\mathbb{E}[\ln(S_1 \cdot x)] : S_0 \cdot x = 1\}$$
is finite. By Lemma 3.14, $x \mapsto \mathbb{E}[\ln(S_1 \cdot x)]$ is strictly concave. Thus problem (6.1) has a unique optimal portfolio. Q.E.D.

The growth optimal portfolio has the nice property that it provides the fastest compounded growth of the capital. By Remark 3.25 (b) it is independent of any risk measures. In the special case that all the risky assets are representing a certain gaming outcome, $\kappa$ is the Kelly allocation in [8]. However, the growth portfolio is seldomly used in investment practice for being too risky. The book [11] edited by MacLean, Thorp, and Ziemba provides an excellent collection of papers with chronological research on this subject. These observations motivated Vince [28] to introduce his *leverage space portfolio* to scale back from the growth optimal portfolio. Recently, [10, 30] further introduce systematical methods to scale back from the growth optimal portfolio by, among other ideas, explicitly accounts for limiting a certain risk measure. The analysis in [10, 30] can be phrased as solving

$$\gamma(\mu) := \inf\{\mathfrak{r}(x) = \widehat{\mathfrak{r}}(\widehat{x}) : \mathbb{E}[\ln(S_1 \cdot x)] \geq \mu, S_0 \cdot x = 1\}, \tag{6.10}$$

where $\mathfrak{r}$ is a risk measure that satisfies conditions (r1) and (r2). Alternatively, to derive the efficient frontier we can also consider

$$\nu(r) := \sup\{\mathbb{E}[\ln(S_1 \cdot x)] : \mathfrak{r}(x) = \widehat{\mathfrak{r}}(\widehat{x}) \leq r, S_0 \cdot x = 1\}, \tag{6.11}$$

Applying Proposition 3.22, Theorem 3.24 and Remark 3.25 to the set of admissible portfolios $A = \{x \in \mathbb{R}^{M+1} : S_0 \cdot x = 1\}$ we derive

**Theorem 6.4.** (Leverage Space Portfolio and Risk Measure) *We assume that the financial market $S_t$ in Definition 2.1 has no nontrivial riskless portfolio and that the risk measure $\mathfrak{r}$ satisfies conditions (r1), (r1n) and (r2). Then*

*(a) problem (6.10) defines $\gamma(\mu) : [\ln(R), \mu_\kappa] \to \mathbb{R}$ as a continuous increasing convex function, where $\mu_\kappa := \mathbb{E}[\ln(S_1 \cdot \kappa)]$ and $\kappa$ is the optimal growth portfolio. Moreover, problem (6.10) has a continuous path of unique solutions $x(\mu)$ that maps the interval $[\ln(R), \mu_\kappa]$ into a curve in the leverage portfolio space $\mathbb{R}^{M+1}$. Finally, $x(\ln(R)) = (1, \widehat{0})^\top$, $x(\mu_\kappa)) = \kappa$, $\gamma(\ln(R)) = \widehat{\mathfrak{r}}(\widehat{0}) = 0$ and $\gamma(\mu_\kappa) = \mathfrak{r}(\kappa)$.*

*(b) problem (6.11) defines $\nu(r) : [0, \mathfrak{r}(\kappa)] \to \mathbb{R}$ as a continuous increasing concave function, where $\kappa$ is the optimal growth portfolio. Moreover, problem (6.11) has a continuous path of unique solutions $y(r)$ that maps the interval $[0, \mathfrak{r}(\kappa)]$ into a curve in the leverage portfolio space $\mathbb{R}^{M+1}$. Finally, $y(0) = (1, \widehat{0})^\top$, $y(\mathfrak{r}(\kappa)) = \kappa$, $\nu(0) = \ln(R)$ and $\nu(\mathfrak{r}(\kappa)) = \mu_\kappa$.*

**Proof.** Note that Assumption 3.15 (a) holds due to Lemma 6.3 and (c2) in Theorem 3.24 is also satisfied. Then (a) follows straight forward from conclusions (a) and (c) in Theorem 3.24 where $\mu_{\max} = \mu_\kappa$ and $r_{\min} = 0$ are finite and attained and (b) follows from conclusions (b) and (c) in Theorem 3.24 with $\mu_{\min} = \ln(R)$ and $r_{\max} = \gamma(\kappa)$.  Q.E.D.

Theorem 6.4 relates the leverage portfolio space theory to the framework setup in Section 3. It becomes clear that each risk measure satisfying conditions (r1), (r1n) and (r2) generates a path in the leverage portfolio space connecting the portfolio of a pure riskless bond to the growth optimal portfolio. Theorem 6.4 also tells us that different risk measures usually correspond to different paths in the portfolio space. Many commonly used risk measures satisfy conditions (r1) and (r2). The curve $x(\mu)$ provides a pathway to reduce risk exposure along the efficient frontier in the risk-expected log utility space. As observed in [10, 30], when investments have only a finite time horizon then there are additional interesting points along the path $x(\mu)$ such as the inflection point and the point that maximizes the return/risk ratio. Both of which provide further landmarks for investors.

Similar to the previous sections we can also consider the related problem of using only portfolios involving risky assets, i.e.,

$$\max_{\widehat{x} \in \mathbb{R}^M} \quad \mathbb{E}[\ln(\widehat{S}_1 \cdot \widehat{x})] \tag{6.12}$$
$$\text{Subject to} \quad \widehat{S}_0 \cdot \widehat{x} = 1.$$

**Theorem 6.5.** (Existence of Solutions) *Suppose that*

$$S_1^i(\omega) > 0, \ \forall \omega \in \Omega, i = 1, \ldots, M. \tag{6.13}$$

*Then problem (6.12) has a solution.*

**Proof.** As in the proof of Theorem 6.4, we can see that Assumption 3.15 (a) holds due to Lemma 6.3. Observe that for $\widehat{x}^* = (1/M, 1/M, \ldots, 1/M)^\top$ we get from (6.13) that $\mathbb{E}[\ln(\widehat{S}_1 \cdot \widehat{x}^*)]$ is finite. Then we can directly apply Theorem 3.24 with $A = \{x \in \mathbb{R}^{M+1} : S_0 \cdot x = 1, x_0 = 0\}$.  Q.E.D.

However, due to the involvement of the log utility function, the relative location of efficient frontiers (6.11) of (6.1) and (6.12) may have several different configurations. The following is an example.

**Example 6.6.** *Let $M = 1$. Consider a sample space $\Omega = \{0, 1\}$ with probability $P(0) = 0.45$ and $P(1) = 0.55$ and a financial market involving a riskless bond with $R = 1$ and one risky asset specified by $S_0^1 = 1$, $S_1^1(0) = 0.5$ and $S_1^1(1) = 1 + \alpha$ with $\alpha > 9/22$ so that $\mathbb{E}[S_1^1] > S_0^1$. Use the risk measure $\mathfrak{r}_1(x_0, x_1) = |x_1|$ (which is an approximation of the drawdown cf. [30]). Then it is easy to calculate that the efficient frontier (6.11) of (6.1) is*

$$\nu(r) = 0.55 \ln(1 + \alpha r) + 0.45 \ln(1 - 0.5r), r \in [0, r_{\max}^\alpha], \tag{6.14}$$

*where $r_{\max}^\alpha = (22\alpha - 9)/20\alpha$. On the other hand the efficient frontier of (6.12) is a single point $\{(1, \nu(1))\}$ where $\nu(1) = 0.55 \ln(1 + \alpha) - 0.45 \ln(2)\}$.*

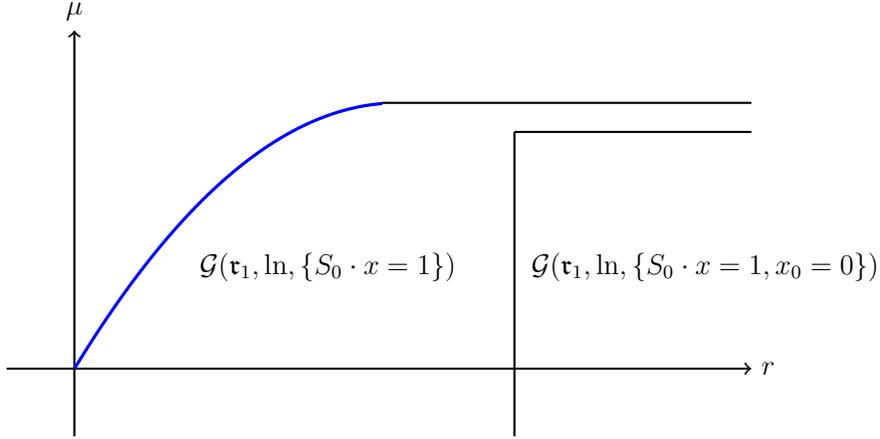

Figure 7: Separated efficient frontiers

When $\alpha \in (9/22, 9/2)$ the two efficient frontiers (6.11) of (6.1) and (6.12) have no common points (see Figure 7). However, when $\alpha \geq 9/2$, $\mathcal{G}_{eff}(\mathfrak{r}_1, \ln, S_0 \cdot x = 1, x_0 = 0) \subset \mathcal{G}_{eff}(\mathfrak{r}_1, \ln, S_0 \cdot x = 1)$ (see Figure 8). In particular, when $\alpha = 9/2$, $\mathcal{G}_{eff}(\mathfrak{r}_1, \ln, S_0 \cdot x = 1, x_0 = 0)$ coincide with the point on $\mathcal{G}_{eff}(\mathfrak{r}_1, \ln, S_0 \cdot x = 1)$ corresponding to the growth optimal portfolio as illustrated in Figure 9.

In fact, a far more common restriction to the set of admissible portfolios are limits of risk. For this example if, for instance, we restrict the risk by $\mathfrak{r}_1(x) \leq 0.5$ then we will create a shared efficient frontier of (6.1) with that of (6.11) where $\mathfrak{r}$ is a priori restricted (see Figure 10).

**Remark 6.7.** (Efficiency Index) *Although the growth optimal portfolio is usually not implemented as an investment strategy, the maximum utility $\mu_{\max}$ corresponding to the growth optimal portfolio $\kappa$, empirically estimated using historical performance data, can be used as a measure to compare different investment strategies. This is proposed in [31] and called the efficiency index. When the only risky asset is the payoff of a game with two outcomes following a given playing strategy, the efficiency coefficient coincides with Shannon's information rate (see [8, 21, 31]). In this sense, the efficiency index gauges the useful information contained in the investment strategy it measures.*

Also related to the growth optimal portfolio theory is the fundamental theorem of asset pricing (FTAP). FTAP characterizes the no arbitrage condition with the existence of a martingale measure, which is defined below.

**Definition 6.8.** (Equivalent Martingale Measure) *We say that $Q$ is an equivalent martingale measure (EMM) for the financial market $S_t$ on a probability space $(\Omega, 2^\Omega, P)$ provided that $Q$ is a probability measure such that, for any $\omega \in \Omega$, $Q(\omega) \neq 0$ if and only if $P(\omega) \neq 0$, and*
$$\mathbb{E}^Q[S_1] = RS_0.$$

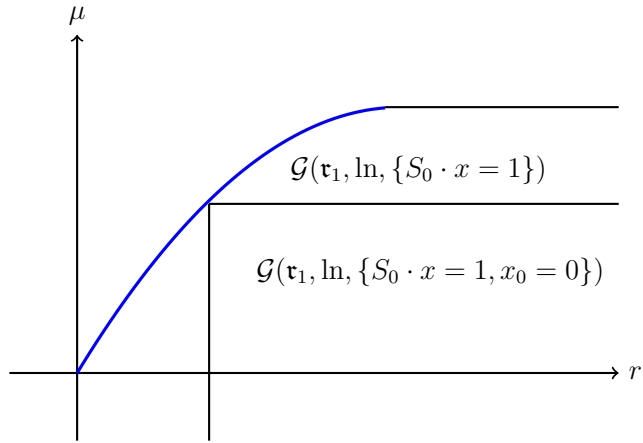

Figure 8: Touching efficient frontiers

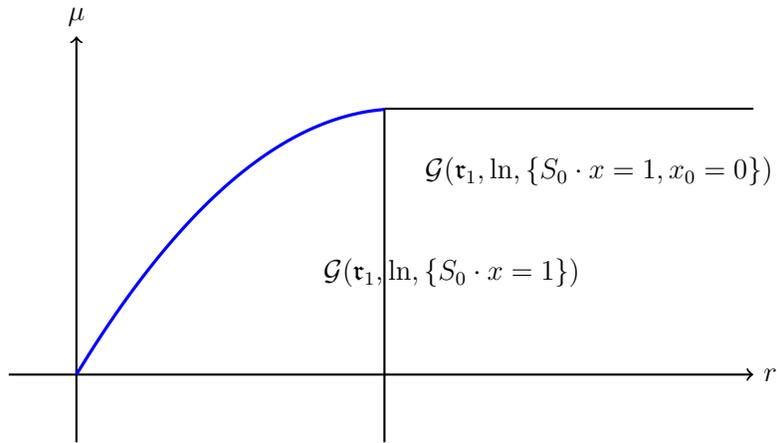

Figure 9: Touching efficient frontiers at growth optimal

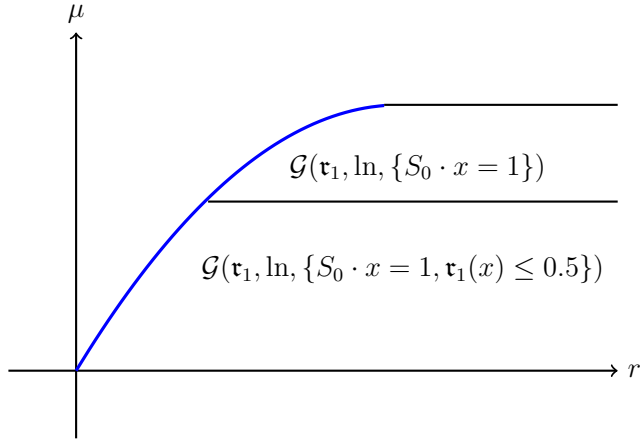

Figure 10: Shared efficient frontiers

We will relate the fundamental theorem of asset pricing to the following general utility optimization problem

$$\max_{x \in \mathbb{R}^{M+1}} \quad \mathbb{E}[u(S_1 \cdot x)] \tag{6.15}$$
$$\text{Subject to} \quad S_0 \cdot x = 1.$$

First we observe that when a utility function $u$ satisfies condition (u4) we can also characterize the no arbitrage condition in terms of the supremum of the expected utility.

**Theorem 6.9.** (Characterization of No Arbitrage) *Suppose that the financial market $S_t$ of Definition 2.1 has no nontrivial portfolio equivalent to the bond. Let $u$ be a utility function satisfying conditions (u3) and (u4) in Assumption 3.3. Then $S_t$ has no arbitrage if and only if*

$$\sup_{x \in \mathbb{R}^{M+1}} \{\mathbb{E}[u(S_1 \cdot x)] : S_0 \cdot x = 1\} < +\infty.$$

**Proof.** Note that

$$\{\mathbb{E}[u(S_1 \cdot x)] : S_0 \cdot x = 1\} = \{\mathbb{E}[u(R + (\widehat{S}_1 - R\widehat{S}_0) \cdot \widehat{x})] : \widehat{x} \in \mathbb{R}^M\}.$$

We can easily verify that when a utility function $u$ satisfies condition (u4) and there exists an arbitrage portfolio then

$$\sup_{\widehat{x} \in \mathbb{R}^M} \{\mathbb{E}[u(R + (\widehat{S}_1 - R\widehat{S}_0) \cdot \widehat{x})]\} = \infty.$$

On the other hand, by Proposition 3.7 when $S_t$ has no nontrivial portfolio equivalent to the bond and no arbitrage implies that $S_t$ has no nontrivial riskless portfolio. By Lemma 6.3, $\{x \in \mathbb{R}^{M+1} : \mathbb{E}[u(S_1 \cdot x)] \geq \mu, S_0 \cdot x = 1\}$ is compact. Thus,

$$\sup_{x \in \mathbb{R}^{M+1}} \{\mathbb{E}[u(S_1 \cdot x)] : S_0 \cdot x = 1\} < +\infty.$$

**Theorem 6.10.** (Fundamental Theorem of Asset Pricing) *Suppose that the financial market $S_t$ of Definition 2.1 has no nontrivial portfolio equivalent to the bond. Let u be a utility function that satisfies properties (u1), (u2s), (u3) and (u4) in Assumption 3.3. Then the following assertions are equivalent:*

(i) *The financial market $S_t$ in Definition 2.1 has no arbitrage.*

(ii) *The optimal value of the portfolio utility optimization problem (6.15) is finite and attained.*

(iii) *There is an equivalent martingale measure for the financial market $S_t$ proportional to a subgradient of $-u$ at the optimal solution of (6.15).*

**Proof.** Observe that (i) equivalent to (ii) is already derived in Theorem 6.9.
To prove (ii) implies (iii) we rewrite the utility optimization problem (6.15) as

$$\max_{y} \quad \mathbb{E}[u(y)] \tag{6.16}$$
$$\text{subject to} \quad R + (S_1(\omega) - RS_0) \cdot x - y(\omega) = 0, \text{ for all } \omega \in \Omega.$$

Assume that $(\bar{x}, \bar{y})$ is the solution to (6.16). Then there exist Lagrange multipliers $\lambda(\omega)P(\omega)$, $\omega \in \Omega$ such that the Lagrangian

$$L((x,y),\lambda) = \mathbb{E}[u(y) + \lambda(R + (S_1 - RS_0) \cdot x - y)]. \tag{6.17}$$

attains an unconstrained maximum at $(\bar{x}, \bar{y})$. Thus, the convex function $-L$ attains an unconstrained minimum at $(\bar{x}, \bar{y})$. It follows that

$$-\lambda(\omega) \in \partial(-u)(\bar{y}(\omega)) \tag{6.18}$$

(so that $\lambda(\omega) > 0$ by (u1) and (u2s)) and

$$\mathbb{E}[\lambda(S_1 - RS_0)] = 0. \tag{6.19}$$

It follows that $Q = (\lambda/\mathbb{E}[\lambda])P$ is an equivalent martingale measure. This process is reversible. Q.E.D.

# 7 Conclusion and Open Problems

Following the pioneering idea of Markowitz to trade-off the expected return and standard deviation of a portfolio, we consider a general framework to efficiently trade-off between a concave expected utility and a convex risk measure for portfolios. Under reasonable assumptions we show that (i) the efficient frontier in such a trade-off is a convex curve in the expected utility-risk space, (ii) the optimal portfolio corresponding to each level of the expected utility is unique and (iii) the optimal portfolios continuously depend on the level of the expected utility. Moreover, we provide an alternative treatment of the results in [20] showing that the one fund theorem (Theorem 4.5) holds in the trade-off between a deviation measure and the expected return (Theorem 5.1) and construct a counter-example illustrating that the two fund theorem (Theorem 4.2) fails in such a general setting. Furthermore, the efficiency curve in the leverage space is supposedly an economic way to scale back risk from the growth optimal portfolio (Theorem 6.4).

This general framework unifies a group of well known portfolio theories. They are Markowitz portfolio theory, capital asset pricing model, the growth optimal portfolio theory, and the leverage portfolio theory. It also extends these portfolio theories to more general settings.

The new framework also leads to many questions of practical significance worthy further explorations. For example, quantities related to portfolio theories such as the Sharpe ratio and efficiency index can be used to measure investment performances. What other performance measurements can be derived using the general framework in Section 3? Portfolio theory can also inform us about pricing mechanisms such as those discussed in the capital asset pricing model and the fundamental theorem of asset pricing. What additional pricing tools can be derived from our general framework?

Clearly, for the purpose of applications we need to focus on certain special cases. Drawdown related risk measures coupled with the log utility attracts much attention in practice. In Part II of this series [14] several drawdown related risk measures are constructed and analyzed. We will conduct a related case study in the third part of this series [3].